\documentclass[12pt,preprint]{aastex}
\newcommand{\bdv}[1]{\mbox{\boldmath$#1$}}

\def\au{{\rm AU}} 
 
\def\kms{{\rm km}\,{\rm s}^{-1}}
\def\masyr{{\rm mas}\,{\rm yr}^{-1}}
\def\kpc{{\rm kpc}}
\def\mas{{\rm mas}}

\def\muas{\mu{\rm as}}

\def\pc{{\rm pc}}

\def\rel{{\rm rel}}

\def\eff{{\rm eff}}

\def\hel{{\rm hel}}

\def\base{{\rm base}}
\def\e{{\rm E}}
\def\bpi{{\bdv\pi}}
\def\bmu{{\bdv\mu}}

\def\bgamma{{\bdv\gamma}}

\def\bv{{\bf v}}

\begin{document}
\title{KMT-2018-BLG-1292: A Super-Jovian Microlens Planet in the Galactic Plane}

\author{\textsc{
Yoon-Hyun Ryu$^{1}$, 
Maria Gabriela Navarro$^{2,3,4}$,
Andrew Gould$^{5,6}$, 
\and
Michael D.Albrow$^{7}$, 
Sun-Ju Chung$^{1,8}$, 
Cheongho Han$^{9}$, 
Kyu-Ha Hwang$^{1}$, 
Youn Kil Jung$^{1}$, 
In-Gu Shin$^{1}$, 
Yossi Shvartzvald$^{10}$, 
Jennifer C. Yee$^{11}$, 
Weicheng Zang$^{12}$,
Sang-Mok Cha$^{1,13}$, 
Dong-Jin Kim$^{1}$,
Hyoun-Woo Kim$^{1}$, 
Seung-Lee Kim$^{1,8}$, 
Chung-Uk Lee$^{1,8}$, 
Dong-Joo Lee$^{1}$,
Yongseok Lee$^{1,13}$, 
Byeong-Gon Park$^{1,8}$, 
Richard W. Pogge$^{6}$\\
(KMTNet Collaboration)\\
Dante Minniti$^{2,4,14}$, 
Roberto K.\ Saito$^{15}$,
Javier Alonso-Garc{\'i}a$^{4,16}$\\
(VVV Collaboration)\\
\and
Matthew T.\ Penny$^{6}$
} }

%----------------------------------------------------------------
\affil{$^{1}$Korea Astronomy and Space Science Institute, Daejon
34055, Republic of Korea}

\affil{$^{2}$Departamento de Ciencias F{\'i}sicas, Facultad de Ciencias 
Exactas,  Universidad Andr\'es Bello, Av. Fernandez Concha 700, Las Condes, 
Santiago, Chile}

\affil{$^{3}$Dipartimento di Fisica, Unversit\`a degli Sutdi di 
Roma ``La Sapeinza'', P.le Aido Moro, 2, 100185 Rome, Italy}

\affil{$^{4}$Instituto Milenio de Astrof{\'i}sica, Av. Vicuna Mackenna 4860, 
782-0436, Santiago, Chile}

\affil{$^{5}$Max-Planck-Institute for Astronomy, K\"{o}nigstuhl 17,
69117 Heidelberg, Germany}

\affil{$^{6}$Department of Astronomy, Ohio State University, 140 W.
18th Ave., Columbus, OH 43210, USA}

\affil{$^{7}$University of Canterbury, Department of Physics and
Astronomy, Private Bag 4800, Christchurch 8020, New Zealand}

\affil{$^{8}$Korea University of Science and Technology, Korea, 
(UST), 217 Gajeong-ro, Yuseong-gu, Daejeon, 34113, Republic of Korea}

\affil{$^{9}$Department of Physics, Chungbuk National University,
Cheongju 28644, Republic of Korea}

\affil{$^{10}$IPAC, Mail Code 100-22, Caltech, 1200 E. California Blvd., 
Pasadena, CA 91125, USA}

\affil{$^{11}$ Center for Astrophysics $|$ Harvard \& Smithsonian, 60 Garden
St., Cambridge, MA 02138, USA}

\affil{$^{12}$Physics Department and Tsinghua Centre for
Astrophysics, Tsinghua University, Beijing 100084, China}

\affil{$^{13}$School of Space Research, Kyung Hee University,
Yongin, Kyeonggi 17104, Republic of Korea}

\affil{$^{14}$Vatican Observatory, V00120 Vatican City State, Italy}

\affil{$^{15}$Departamento de F\'{i}sica, Universidade Federal de Santa 
Catarina, 88.040-908, Florian\'opolis, Brazil}

\affil{$^{16}$Centro de Astronom\'{i}a (CITEVA), Universidad de Antofagasta, 
Av. Angamos 601, Antofagasta, Chile}

%\affil{$^{9}$NASA Postdoctoral Program Fellow}

\begin{abstract}
We report the discovery of KMT-2018-BLG-1292Lb, a super-Jovian 
$M_{\rm planet} = 4.5\pm 1.3\,M_J$ planet orbiting an F or G dwarf 
$M_{\rm host} = 1.5\pm 0.4\,M_\odot$, which
lies physically within ${\cal O}(10\,\pc)$ 
of the Galactic plane.  The source star is a heavily extincted $A_I\sim 5.2$
luminous giant that has the lowest Galactic latitude, $b=-0.28^\circ$, of
any planetary microlensing event.  The relatively blue blended light 
is almost certainly either the host or its binary companion, with 
the first explanation being substantially more likely.  This blend
dominates the light at $I$ band and completely dominates at $R$ and
$V$ bands.  Hence, the lens system can be probed by follow-up
observations immediately, i.e., long before the lens system and 
the source separate due to their relative proper motion.  The system
is well characterized despite the low cadence 
$\Gamma=0.15$--$0.20\,{\rm hr^{-1}}$ of observations and short viewing
windows near the end of the bulge season.  This suggests that
optical microlensing planet searches can be extended to the Galactic
plane at relatively modest cost.

\end{abstract}


\section{{Introduction}
\label{sec:intro}}

As a rule, optical microlensing searches heavily disfavor regions of
high extinction and, as a result, systematically avoid the
Galactic plane.  For example, prior to the start of OGLE-IV
(the fourth phase of the Optical Gravitational Lensing Experiment,
\citealt{ogleref}) in 2010, all but a small fraction of Galactic-bulge
microlensing observations were restricted to the southern bulge
despite the fact that the stellar content of the lines of sight
toward the northern and southern bulge are extremely similar.
With its larger format camera, OGLE-IV began systematically covering
the northern bulge, but mainly at very low cadence.  Hence, it remained
the case that the great majority of observations were
toward the southern bulge.

However, \citet{poleski16} showed that the microlensing event rate
is basically proportional to the product of the surface density of
clump stars and the surface density of stars below some magnitude limit
(in the principal survey band),
e.g., $I<20$; the two numbers being proxies for the column densities 
of lenses and sources, respectively\footnote{His formula, derived
from a fit to OGLE data, is actually slightly more complicated.}.
Guided in part by this work, the Korea Microlensing Telescope Network
(KMTNet, \citealt{kmtnet}) devised an observing strategy that much
more heavily favored the northern bulge, which accounts for about
37\% of the area covered and 24\% of all the observations.  Nevertheless,
even with this more flexible attitude toward high-extinction fields,
KMTNet still followed previous practice in systematically avoiding
the Galactic plane. See Figure~12 of \citet{eventfinder}.

Indeed, there is an additional reason for avoiding fields with high or
very high extinction.  That is, even if the high stellar-lens column
densities near the plane partially compensate for the lower column
density of sources, it remains the case that events, particularly
planetary and binary events, in very high extinction fields are more 
difficult to interpret.  Very often these events have 
caustic crossings from which one can usually measure
$\rho=\theta_*/\theta_\e$, i.e., the ratio of the angular radius of the source
to the Einstein radius.  Then, one can usually determine $\theta_*$
from the offset of the source relative to the red clump in color and
magnitude \citep{ob03262}.  However, the color measurement required
for this technique is only possible if the event is detected in a second
band, which is usually $V$ band in most microlensing surveys.  But
$V$-band observations rarely yield usable results in very high-extinction
fields.  Hence, one must either take special measures to observe the
event in a redder band (e.g., $H$) or one must estimate $\theta_*$ without
benefit of a color measurement, which inevitably substantially increases
the error in $\theta_*$ (and so $\theta_\e = \theta_*/\rho$).

As a result of the almost complete absence of optical microlensing
observations toward the Galactic plane, there is essentially no
experience with how these theoretical concerns translate into
practical difficulties, and similarly no practical approaches to
overcoming these difficulties.  This is unfortunate because the
Galactic plane could potentially provide important complementary
information to more standard fields in terms of understanding the
microlensing event rate and Galactic distribution of planets.

While this shortcoming is widely recognized, the main orientation
of researchers in the field has been to await infrared microlensing
surveys.  \citet{gould95} advocated a 
``K-band microlensing [survey] of the inner galaxy''.   Although his
focus was on regions projected close to the Galactic center, the
same approach could be applied to any high-extinction region, in
particular the Galactic plane.  In fact, PRIME, a 1.8m field telescope 
with 1.3 deg$^2$ camera to be installed
at SAAO in South Africa, will be the first to conduct a completely
dedicated IR microlensing survey (T. Sumi 2019, private communication), 
While the exact survey strategy has not yet been decided, PRIME 
will certainly focus on heavily
extincted regions toward the inner Galaxy.  The 
VISTA Variables in the Via Lactea (VVV; \citealt{vvv-survey1,vvv-survey2})
Microlensing Survey
\citep{vvv-ulens1,vvv-ulens2} has already conducted wide-field IR observations 
covering a
$(20.4^\circ \times 1.1^\circ)$ rectangle of the Galactic plane 
spanning 2010-2015.  They discovered 630 microlensing events.
However, given their low cadence (ranging from 73 to 104 epochs over
6 years), they were not sensitive to planetary deviations.  In addition,
\citet{vvv-ulens3} used VVV near-IR
photometry to search for microlensing events in fields along the
Galactic minor axis, ranging from $b = -3.7$ to $b = 4$, 
covering a total area of $\sim 11.5\,{\rm deg}^2$. They found $N=238$
new microlensing events in total, $N=74$ of which have bulge red clump
(RC) giant sources.  They found a strong increase of the number of
microlensing events with Galactic latitude toward the plane, both in
the total number of events and in the RC subsample, in particular,
an order of magnitude more events at $b = 0$
than at $|b| = 2$ along the Galactic minor axis.  This
gradient is much steeper than predicted by models that had
in principle been tuned to explain the observations from the optical
surveys farther from the plane.

\citet{ukirt-5events} conducted a survey of 
high-extinction microlensing fields (Figure~1 of \citealt{ukirt-5events}
and Figure~1 of \citealt{ub17001}), 
which had substantially higher
cadence despite the relatively short viewing window from the 3.8m UKIRT
telescope in Hawaii.  This yielded the first infrared detection of 
a microlensing planet, UKIRT-2017-BLG-001Lb, which lies projected 
just $0.33^\circ$ from the Galactic plane and $0.35^\circ$ from the Galactic
center \citep{ub17001}. Both values were by far the smallest for
any microlensing planet up to that point.  
They estimated the extinction at $A_K=1.68$,
which corresponds approximately to $A_I\simeq 7 A_K = 11.8$.

This high extinction value might lead one to think that such planets 
are beyond the reach of optical surveys.  In fact, KMTNet routinely
monitors substantial areas of very high extinction simply because its cameras
are so large that these are ``inadvertently'' covered while observing
neighboring regions of lower extinction and high stellar density.
For example, KMT-2018-BLG-0073\footnote{http://kmtnet.kasi.re.kr/ulens/event/2018/view.php?event=KMT-2018-BLG-0073} 
lies at $(l,b)=(+2.32,+0.27)$ and has $A_K=1.3$.
This raises the possibility that optical surveys could in fact probe
very high extinction regions as well, albeit restricted to monitoring
exceptionally luminous sources or very highly magnified events.

Here we report the discovery of the planet KMT-2018-BLG-1292Lb, which
at Galactic coordinates $(l,b)=(-5.23,-0.28)$ is the closest to the
Galactic plane of any microlensing planet to date.  The planetary
perturbation is well characterized despite the fact that it occurred
near the end of the season when it could be observed only about three
hours per night from each site and that it lies in KMTNet's lowest
cadence field.  Thus, this detection in the face of these 
moderately adverse conditions suggests that
optical surveys could contribute to the study of Galactic-plane
planetary microlensing at relatively modest cost.

% VVV
%  J     H     K
% 14.39 13.19 12.81 

% glimpse
%  K     3.6   4.5
% 12.32 11.90 11.73

% GAIA
%  plx      eplx pmra  epmra  pmdec epmdec gmag   egmag  off x  off y
% -0.3901 0.5790 0.873 1.116 -2.979 0.809 19.3414 0.0061 -53      +6
%           jd    7022   vs    jd 1039.48       2mass   +233    +473
% (-286,467)/16.38 = (17.5,29.5)

\section{{Observations}
\label{sec:obs}}

KMT-2018-BLG-1292 is at (RA,Dec) = (17:33:42.62,$-33$:31:14.41)
corresponding to $(l,b)=(-5.23,-0.28)$.  It was discovered by
applying the KMTNet event-finder algorithm \citep{eventfinder}
to the full-season of 2018 KMTNet data, which were taken from three
identical 1.6m telescopes equipped with $(2^\circ \times 2^\circ)$
cameras in Chile (KMTC), South Africa (KMTS), and Australia (KMTA).
The event lies in KMT field BLG13, which was observed in the $I$ band
at cadences of 
$\Gamma=0.2\,{\rm hr}^{-1}$ from KMTC and 
$\Gamma=0.15\,{\rm hr}^{-1}$ from KMTS and KMTA.
One out of every ten $I$-band observations was matched by an observation
in the $V$ band.  However, the $V$-band light curve is not useful due to high
extinction.

The event was initially classified as ``clear microlensing'' based on
the relatively rough DIA pipeline photometry \citep{alard98,wozniak2000},
but planetary features were not obvious.
The possibly planetary anomaly was noted on 5 January 2019, when 
the data were routinely re-reduced using the KMTNet pySIS \citep{albrow09} 
pipeline as part of the event-verification process.  
The first modeling was carried almost immediately, on 8 January 2019.
This confirmed the planetary nature, thus triggering final
tender-loving care (TLC) reductions.  But, in addition, it also made clear 
that the event
might still be ongoing after the bulge had passed behind the Sun.

This led KMTNet to take two measures to obtain additional data.
First, KMTNet began observing BLG13 from KMTC on 2 February, 
which was 17 days before the start of its general bulge observations.
This was made possible by the fact that KMT-2018-BLG-1292 lies near
the western edge of the bulge fields and so can be observed earlier
in the season than most fields, given the pointing restrictions due
to the telescope design.  Second, KMTNet contacted C.~Kochanek
for special permission to obtain nine epochs of observations 
(17 pointings) from 31 January 2019 
to 8 February 2019 on the dual channel (optical/infrared)
ANDICAM camera \citep{depoy03} on the 1.3m SMARTS
telescope in Chile.  The primary objective of these observations
was to obtain $H$-band data, which could yield an $I-H$ color,
provided that the event remained magnified at these late dates.
As mentioned above, it was already realized that the KMT $V$-band
data would not yield useful source-color information.

However, because the source turned out to be a low-amplitude
variable (see Section~\ref{sec:var}) while the magnification 
at the first ANDICAM $H$-band observation was low, $A\sim 1.1$, 
the $(I-H)$ color measurement from these data was significantly
impacted by systematic uncertainties.  Fortunately, the VVVX survey
\citep{vvvx-survey}
obtained seven $K_s$-band data points on the rising part of the
light curve, including three with magnifications $A=1.47$--1.58.
While these are, of course, also affected by systematics from
source variability, the impact is a factor $\sim 5$ times smaller.
Hence, in the end, we use these VVV survey data to measure the source
color.

%Reductions for the source-color analysis were done using DoPhot \citep{dophot}.

\section{{Light Curve Analysis}
\label{sec:anal}}

\subsection{{Source and Baseline Variability}
\label{sec:var}}

The light curve exhibits low-level (few percent) variability, including
roughly periodic variations with period $P\sim 13\,$days.
This level of variation is far too small to have important implications 
for deriving basic model parameters, but could in principle
affect subtle higher-order 
effects, in particular the microlens parallax.  
For clarity of exposition, we therefore
initially ignore this variability when exploring static models
(Section~\ref{sec:static}), and then use these to frame the investigation
of the variability.  We then account for its impact on
the microlensing parameters (and their uncertainties) after introducing
higher-order effects into the modeling in Section~\ref{sec:par}.

\subsection{{Static Model}
\label{sec:static}}

\begin{figure}
\plotone{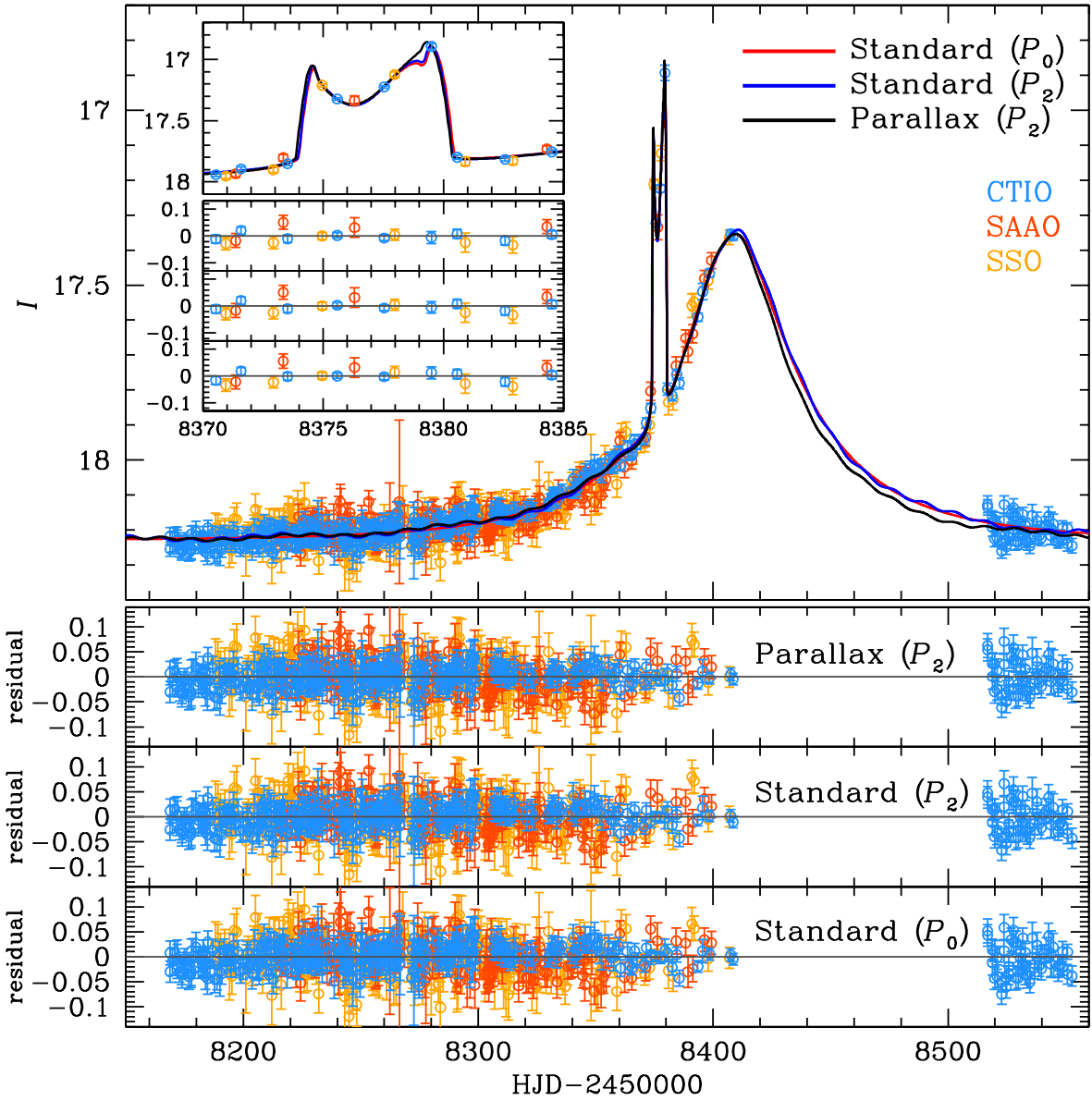}
\caption{KMT data and best-fit model for KMT-2018-BLG-1292.  The lower
three panels show the residuals from the final parallax model, a standard
model that includes two periodic wave forms, and a standard model without
additional wave forms,
respectively.  The inset shows a zoom of the caustic region.
Note that although the source spent six days transiting the caustic,
there are only six data points from all three KMT observatories
combined.  This is partly because the event lies in a low-cadence
field and partly because the anomaly occurred very near the end of
the season, when the bulge is visible for only a few hours per night.
}
\label{fig:lc}
\end{figure}

Figure~\ref{fig:lc} shows the KMT data and best-fit model for 
KMT-2018-BLG-1292.
With the exception of a strong anomaly lasting $\delta t\simeq 6\,$days, 
the 2018 data take the form of the rising half of a standard 
\citet{pac86} single-lens single-source (1L1S) curve.  The
early initiation of 2019 observations, discussed in Section~\ref{sec:obs},
then capture the extreme falling wing of the same Paczy\'nski profile.

We therefore begin by searching for static binary (2L1S) models,
which are characterized by seven non-linear parameters:
$(t_0,u_0,t_\e,q,s,\alpha,\rho)$. 
The first three are the standard 1L1S Paczy\'nski parameters,
i.e., the time of lens-source closest approach, the impact parameter
(in units of the Einstein radius $\theta_\e$), and the Einstein
radius crossing time.  The next three characterize the planet, i.e.,
the planet-host mass ratio, the magnitude of the planet-host
separation (in units of $\theta_\e$), and the orientation of this
separation relative to the lens-source relative proper-motion $\bmu_\rel$.
The last, $\rho\equiv \theta_*/\theta_\e$, is the normalized source
radius.

We first conduct a grid search over $(s,q)$, in which these two
parameters are held fixed while all others are allowed to vary in
a Markov chain Monte Carlo (MCMC).  The Paczy\'nski parameters are seeded
at values derived from a 1L1S fit (with the anomaly removed), and
$\alpha$ is seeded at six values drawn uniformly around the unit circle.
Given the very high extinction
toward this field $A_I\simeq 7 A_K = 5.2$ and the relatively bright
baseline flux $I_{\rm base}\sim 18.2$, the source is very likely to be a
giant.  In view of this, we seed the normalized source radius at $\rho=0.005$.
This procedure yields only one local minimum.  We then
allow all seven parameters to vary and obtain the result shown as
the first model in Table~\ref{tab:ulens}.

\begin{figure}
\plotone{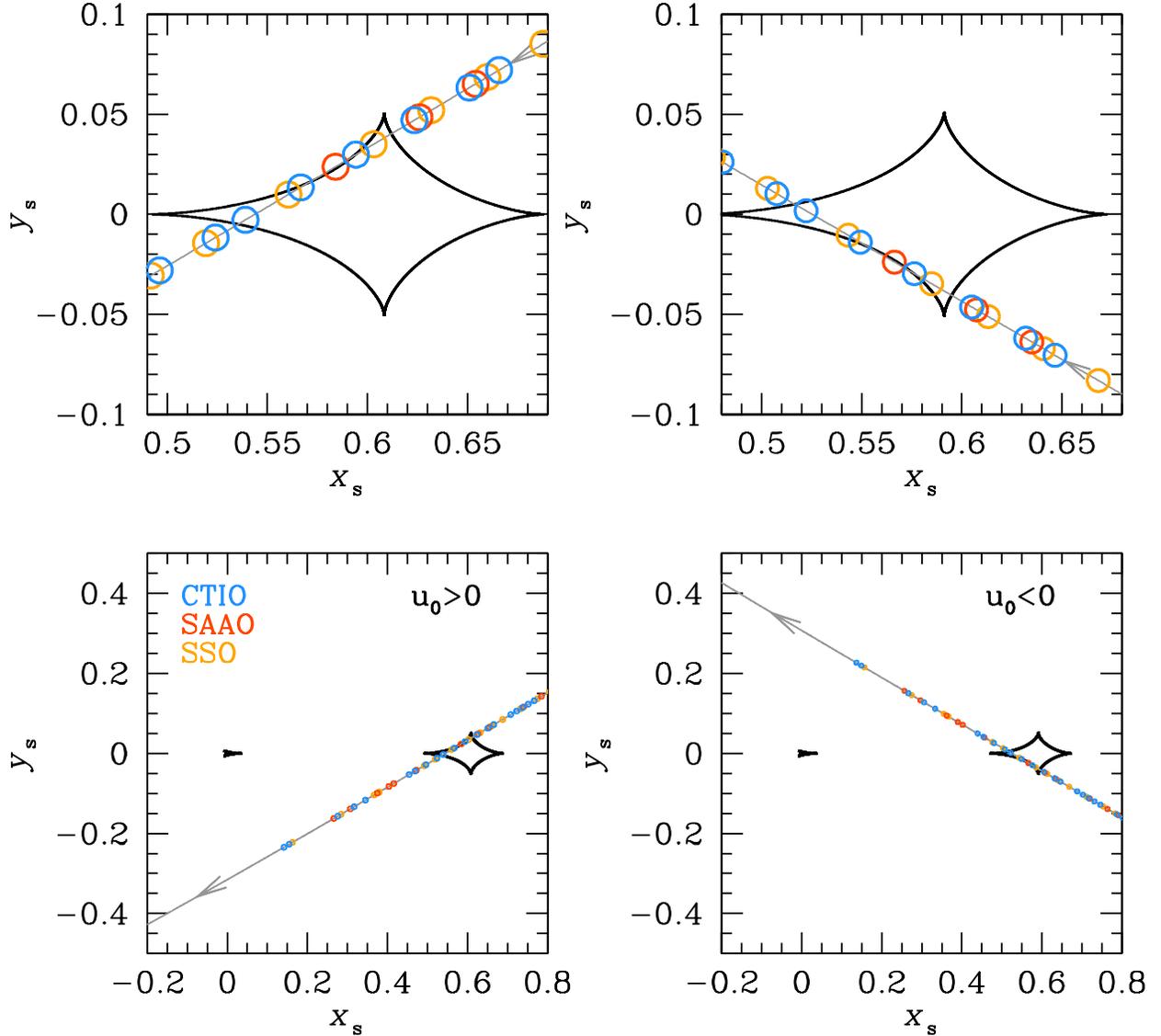}
\caption{Geometry for the two parallax models ($u_0>0$ and $u_0<0$)
of KMT-2018-BLG-1292.  The closed contours show the planetary caustic.
The upper panels are zooms of the regions surrounding these
caustics.  The source size is shown to scale at the epochs of observation,
which are color-coded by observatory.  Note that the source travels along
the edge of the caustic, so that all six data points (spread out over
six days) are affected by the caustic, which enables a reasonably
good measurement
of the normalized source size $\rho=\theta_*/\theta_\e$.  While this
close alignment of the source trajectory with the edge of the caustic
is unusual, the large value of $\rho$ (due to the very large source) 
implies that random trajectories through the caustic would likely 
intersect or closely approach the caustic contour several times.
}
\label{fig:caustic}
\end{figure}

The only somewhat surprising element of this analysis is that $\rho$ 
is measured reasonably well, with $\sim 15\%$ precision.  This is
unexpected because
one does not necessarily expect to measure $\rho$ with such
sparse sampling, roughly one point per day.  However, from the solution,
the source-radius crossing time is $t_*\equiv \rho t_\e = 9.4\,$hrs,
so that the diameter crossing time is almost one day.  Moreover,
as shown by the caustic geometry in Figure~\ref{fig:caustic}, 
the source actually
runs almost tangent to caustic, which means that all six
data points are affected by the caustic (and so finite-source effects).  
Hence, the relatively
good measurement of $\rho$ is partly due to a generic characteristic of
giant-star sources (which in turn are much more likely for optical
microlensing searches in extincted fields) and partly due to a chance
alignment of the source trajectory with the caustic.  We note that
UKIRT-2017-BLG-001 \citep{ub17001} had a similarly good ($\sim 10\%$)
$\rho$ measurement with similar \cal{O}(1 day) 
cadence\footnote{Formally, the cadence was $\Gamma = 3\,{\rm day}^{-1}$ 
compared to an average of $\Gamma \sim 1\,{\rm day}^{-1}$ for 
KMT-2018-BLG-1292.  However, these three points were confined to
a few hours (see Figure~1 of \citealt{ub17001}), so the gaps in the
data were similar.}, and for
similar reasons: large source, whose detection was favored by heavy
extinction, and consequently long $t_*$ ($\sim 16\,$ hrs).

\subsection{{Parallax Models}
\label{sec:par}}

We next attempt to measure the microlens parallax vector
\citep{gould92,gould00},
\begin{equation}
\bpi_\e \equiv {\pi_\rel\over\theta_\e}{\bmu_\rel\over\mu_\rel},
\qquad
\theta_\e^2 \equiv \kappa M\pi_\rel,
\qquad
\kappa \equiv {4G\over c^2\au}\simeq 8.1\,{\mas\over M_\odot}.
\label{eqn:piethetae}
\end{equation}
where, $M$ is the lens mass, $\bmu_\rel$ is the instantaneous geocentric
lens-source relative proper motion, and $\pi_\rel$ is the lens-source
relative parallax.  Because the parallax effect due to Earth's annual
motion is quite subtle, such a measurement can be affected 
by source variability.  Hence we must simultaneously model this 
variability together with the microlens parallax in order to assess 
its impact on both the best estimate and uncertainty of $\bpi_\e$.

\subsubsection{{Significant Parallax Constraints Are Expected}
\label{sec:expect}}

The relatively long timescale, $t_\e\simeq 67\,$days, of the standard
solution in Table~\ref{tab:ulens} suggests that it may be possible to measure
or strongly constrain $\bpi_\e$.  In addition to the relatively
long timescale, the presence of sharply defined peaks (from the 
anomaly) tend to improve microlens parallax measurements \citep{angould}.
Finally, while it would be relatively difficult to measure $\bpi_\e$
from 2018 data alone (because these contain only the rising part of the
light curve), the 2019 data on the extreme falling wing add significant
constraints to this measurement.
We therefore add two parameters to the modeling $(\pi_{\e,N},\pi_{\e,E})$,
i.e., the components of $\bpi_\e$ in equatorial coordinates.

Because parallax effects, which are due to Earth's orbital motion,
can be mimicked in part by orbital motion of the lens system
\citep{mb09387,ob09020}, one should always include lens motion,
at least initially, when incorporating parallax into the fit.
We model this with two parameters, $\bgamma\equiv ((ds/dt)/s,d\alpha/dt)$,
where $ds/dt$ is the instantaneous rate of change in separation and $d\alpha/dt$
is the instantaneous rate of change of the orientation of the binary axis.
Note that all ``instantaneous'' quantities $(\bmu,\bgamma)$ are defined
at time $t=t_0$.  However, we find that these two additional parameters
are not significantly correlated with the parallax and are also not
significantly constrained by the fit.  Hence, we remove them from the fit.

\subsubsection{{Accounting for Variability}
\label{sec:account}}

As mentioned in Section~\ref{sec:var}, the source shows low-level
variations in the standard-model residuals.
We will show in Section~\ref{sec:cmd} that
the source is a luminous red giant, so source variability would not
be unexpected.  These variations
do not significantly affect the static model (and so were ignored up
to this point) but could affect the parallax measurement, which depends
on fairly subtle distortions of the light curve relative to the one
defined by a static geometry.
We therefore simultaneously fit for this variability together
with the nine other non-linear parameters describing the 2L1S parallax solution.
This will allow us, in particular, to determine whether the parallax 
parameters $(\pi_{\e,N},\pi_{\e,E})$ are correlated with the variability 
parameters.  We consider models that incorporate variability into
an ``effective magnification''
\begin{equation}
A_\eff(t) = A(t;t_0,u_0,t_\e,q,s,\alpha,\rho,\pi_{\e,N},\pi_{\e,E})
\biggl[1 + \sum_{i=1}^{N_{\rm per}}a_i\sin\biggl({2\pi t\over P_i} 
+\phi_i \biggr)\biggr],
\label{eqn:var}
\end{equation}
where $(a_i,P_i,\phi_i)$ are the amplitude, period, and phase of each
of the $N_{\rm per}$ wave forms that are included.

We search for initial values of the wave-form parameters by first
applying Equation~(\ref{eqn:var}) to static models with the microlensing
parameters seeded at the best fit non-variation model.  
We set $N_{\rm per}=1$ and find the three
wave-form parameters.  We then set $N_{\rm per}=2$ and seed the previous
$(7+3)=10$ non-linear parameters at the $N_{\rm per}=1$ solution in order to
find the next three.  In principle this procedure could be repeated,
but we find no additional significant periodic variations.

We seeded the first component with $P_1=11\,$days based on our by-eye
estimate of the periodic variations.  Somewhat surprisingly, this
fit converged to $P_1\sim 70\,$days.  Hence, we seeded the second
component again with $P_2=11\,$days, which converged to $P_2 \simeq 13\,$days.
We show this $N_{\rm per}=2$ standard model in Table~\ref{tab:ulens} next
to the model without periodic variation.  
As anticipated in Section~\ref{sec:static}, the introduction of periodic
components has virtually no effect on the standard microlensing parameter
estimates, although the fit is improved by $\Delta\chi^2=27$ for six 
degrees of freedom (dof).

These values served as benchmarks for the next phase of
simultaneously fitting for parallax and periodic variations, in
which the parallax fits could in principle become coupled to
long-term variations.  We seed the $N_{\rm per}=1$ parallax fits
with a variety of periods, but these always converge to $P_1\sim 63\,$days.
We then seed $P_2=13\,$days, which then converges to a similar value.
Adding more wave forms does not significantly improve the fit.

\subsubsection{{Parallax Model Results}
\label{sec:parres}}

Table~\ref{tab:ulens} shows the final results, i.e., 
for nine microlensing parameters plus six periodic-variation parameters.
As usual, we test for the ``ecliptic degeneracy'', which approximately
takes $(u_0,\alpha,\pi_{\e,N})
  \rightarrow -(u_0,\alpha,\pi_{\e,N})$ \citep{ob09020} and
present this solution as well in Table~\ref{tab:ulens}.

In addition, in Table~\ref{tab:evolve}, we show the evolution of key
microlens parameters as additional period terms are introduced.
In fact, neither the microlens parallax nor the other key microlens 
parameters change significantly 
as a result of incorporating periodic variability into the fits.

Because both $\pi_\e$ and $\rho$ are measured, one can infer the
lens mass and lens-source relative parallax via,
\begin{equation}
M = {\theta_\e\over\kappa \pi_\e},
\qquad
\pi_\rel = \theta_\e\pi_\e,
\label{eqn:massdist}
\end{equation}
provided that the angular source size $\theta_*$ 
(and so $\theta_\e = \theta_*/\rho$) can be determined
from the color-magnitude diagram (CMD).

\section{{Color-Magnitude Diagram}
\label{sec:cmd}}

There are two challenges to applying the standard procedure \citep{ob03262}
of putting the source star on an instrumental CMD
in order to determine $\theta_*$.  Both challenges derive from the
fact that the event lies very close to the Galactic plane.  First, the 
extinction is high, which implies that the $V$-band data, which are routinely
taken, will not yield an accurate source color.  
Fortunately, there are $K_s$ data from the VVVX survey taken
when the event was sufficiently magnified to measure the $K_s$ source flux.

\begin{figure}
\plotone{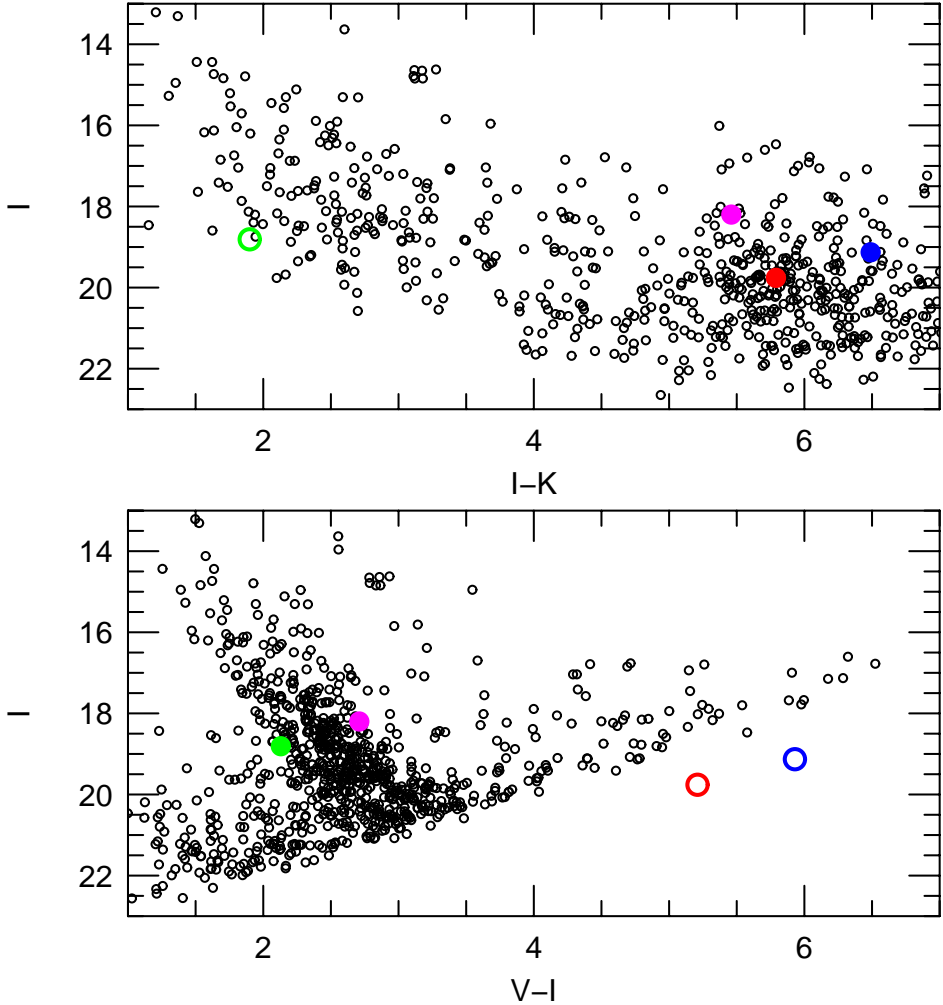}
\caption{Color-magnitude diagrams (CMDs) in $I$ vs.\ $(I-K)$ (upper)
and $I$ vs.\ $(V-I)$ (lower).  The black points are field stars from
a $(2^\prime \times 2^\prime)$ square centered on KMT-2018-BLG-1292.
The large circles are the positions of the source (blue), blend (green),
``baseline object'' (magenta) and clump centroid (red).  The filled
circles are measured, while the open circles are estimated (and shown
for illustration only).  The source (blue) is a luminous and very red giant.
The blend (green) is a foreground main-sequence star, lying in front
of the majority of the dust column toward the Galactic bar.
}
\label{fig:cmd}
\end{figure}

The second issue is more fundamental. The upper panel in 
Figure~\ref{fig:cmd} shows an $I$ versus
$(I-K)$ CMD, where the $I$-band data come from pyDIA reductions of
the field stars within a $2^\prime\times 2^\prime$ square centered on 
the event and the $K$-band data come
from the VVV catalog.  The position of the ``baseline object'' (magenta)
is derived from the field-star photometry of these two surveys, while
the position of the source star (blue) is derived from the $f_S$ 
measurements from the model fit to the light curves.  The position of
the blended light is shown as an open circle because, while its $I$-band
magnitude is measured from the fit, its $K$-band flux is too small to
be reliably determined.  Hence its position is estimated from the $I$ versus
$(V-I)$ CMD, which is described immediately below.
The centroid of the red clump is shown in red.  

The lower panel of Figure~\ref{fig:cmd}
shows the same quantities for the $I$ versus $(V-I)$ CMD.
It is included to facilitate analysis of the properties of the blend,
which is discussed further below.
In this case, the source (blue) and clump centroid (red) are shown
as open symbols because neither can be reliably determined from the
data and so are estimates rather then measurements.

The source lies $\Delta(I-K,I)=(+0.70,-0.63)$ redward and brighter
than the clump.  We first interpret this position under the assumption
that the lens suffers similar extinction as the clump itself.  In this
case, the source is a very red, luminous giant, 
$[(I-K)_0,M_I]\simeq (2.1,-0.7)$, which would explain why it is a 
low-amplitude semi-regular variable.
% 0.245 +- 0.005   0.584 +- 0.0014  V (V-K) log theta_0 = a(V-K) + b
% 0.245*(V-K) - 0.2*V + 0.584 - log(2)
% 0.045*(V-K) - 0.2*K + 0.283
% I = 14.10 (I-K)=1.78 (V-K)=3.85 K=12.32
% theta_* = 10^(0.045*3.85 - 0.2*12.32 + 0.283) = 9.8 uas

Adopting the assumption that the source suffers the same extinction
as the clump, together with the intrinsic clump position 
$[(V-I),I]_{0,\rm cl} = (1.06,14.66)$ from \citet{bensby13} and 
\citet{nataf13}, as well as the color-color relations of \citet{bb88},
we obtain $[(V-K),K]_0 = (3.90,11.87)$.  Then using the
color/surface-brightness relation of \citet{groenewegen04}
\begin{equation}
\log (\theta_*/\muas) = 3.286 - 0.2\,K_0 + 0.039(V-K)_0,
\label{eqn:csb}
\end{equation}
we obtain
\begin{equation}
\theta_* = 11.59\pm 1.00\,\muas.
\label{eqn:thetastareval}
\end{equation}

The error bar in Equation~(\ref{eqn:thetastareval}) is determined as follows.
First, while the formal error $\Delta(I-K)$ (from fitting the $I$ and $K$
light curves to the model and centroiding the clump) is only $\sim 0.05\,$mag,
we assign a total error $\sigma[\Delta(I-K)]= 0.11\,$mag (i.e., adding
0.1 mag in quadrature).  We do so because the source is variable,
and this variation may have a different phase and amplitude in $I$
(where it is measured) than $K$.  Hence, we determine $I-K$ by fitting
both light curves to a standard model without periodic wave-forms
and account for the unknown form of the variation with this error term.
This error directly propagates to errors of 0.28 mag in $(V-K)_0$ and 0.11 mag
in $K_0$, which are perfectly anti-correlated, and so add constructively
via Equation~(\ref{eqn:thetastareval})  to
$0.2\times 0.11 + 0.039\times 0.28 = 0.329\,$dex.  Finally, there
is a statistically independent error in $\Delta I$ of 0.09 mag, which
comes from a 0.07 mag error in centroiding the clump and a 0.05 mag
error from fitting the model.  This yields an additional error in
Equation~(\ref{eqn:thetastareval})  of $0.2\times 0.09= 0.018\,$dex, which
is added in quadrature to obtain the final result.

We consider the assumption underlying Equation~(\ref{eqn:thetastareval})
that the source suffers the same extinction
as the clump to be plausible because there is a well-defined
clump, meaning that there is a strong overdensity of stars at the bar.
Hence, it is quite reasonable that the source would lie in this
overdensity.  However, because the line of sight passes through the
bar only about 45 pc below the Galactic plane, it is also possible that the
source lies in front of, or behind, the bar.  For example, 
the source star for UKIRT-2017-BLG-001Lb, the only other 
microlensing planet that was discovered so close to the Galactic plane,
was found to lie in the far disk \citep{ub17001}.  From the standpoint of
determining $\theta_*$, the distance to the source does not enter directly
because only the apparent magnitude and color enter into 
Equation~(\ref{eqn:csb}).  But the distance does enter indirectly
because if the source lies farther or closer than the clump,
then it suffers more or less extinction.  In most microlensing events
this issue is not important because the line of sight usually intersects
the bulge well above (or below) the dust layer.  We can parameterize the
extra dust (or dust shortfall) relative to the clump by $\Delta A_K$.
Then, from Equation~(\ref{eqn:csb}), the inferred change in $\theta_*$
for a given excess dust column is
\begin{equation}
{\Delta\log\theta_*\over \Delta A_K} = 
0.2\biggl(0.195{E(V-K)\over A_K}-1\biggr) \rightarrow 0.23,
\label{eqn:dustderiv}
\end{equation}
where we have adopted $E(V-K)=11\,A_K$.

The dust column to the clump has $A_K = 0.75$.  The source cannot
lie in front of substantially less dust than the clump
because then it would be intrinsically both much redder and much
less luminous than we derived above for the color and absolute magnitude.
For example, if $\Delta A_K= -0.1$ and the source were at $D_S=6\,\kpc$
then, $[(I-K)_0,M_I]\rightarrow (2.7,+0.9)$.  Such low luminosity extremely
red giants are very rare.

By the same token, if $\Delta A_K = +0.1$ and $D_S=11\,\kpc$, then
$[(I-K)_0,M_I]\rightarrow (1.5,-1.8)$.  This is a marginally
plausible combination, although higher values of $A_K$ would imply
giants that are bluer than the clump but several magnitudes brighter.
We adopt a $1\,\sigma$ uncertainty in $\sigma(A_K) = 0.05$, and
hence a fractional error $\sigma(\ln\theta_*) = 0.05\cdot 0.23\,\ln 10 = 2.6\%$.
This uncertainty is actually small compared to the 8.6\% error in
Equation~(\ref{eqn:thetastareval}).  Finally we adopt an error of 9.0\%
by adding these two errors in quadrature.  (We will provide some evidence
in Section~\ref{sec:pietest} that the source is actually in the bar.)

% mu_L = (0.8 -3.0) +-(1.1,0.8) mu_rel = (8.0,0.1)  
% mu_s =   (-7.2,-3.1) +-(1.8,0.8)
% mu_s_exp (-3.3,-5.7) +-(2.5,2.5)  diff (3.9,2.6)+-(3.0,2.6)
% chi2 = 2.69 for 2 dof
Combining the value of $\theta_*$ from Equation~(\ref{eqn:thetastareval})
with the average of the two virtually identical values of $\rho$ in 
Table~\ref{tab:ulens} (but using the larger error), we obtain
% effectively 6.510 +- 1.135 e-3 (because must use larger error)=17.4%
% total error theta* + rho is 19.8%
\begin{equation}
\theta_\e = {\theta_*\over \rho} = 1.72\pm 0.34\,\mas
\qquad
\mu_\rel = {\theta_\e\over t_\e} = 10.7 \pm 2.0\,\masyr
\label{eqn:thetaemu}
\end{equation}
Together with the parallax measurement $\pi_\e\sim 0.125$, 
this result for $\theta_\e$ implies that
the lens mass and relative parallax are $M\sim 1.7\,M_\odot$
and $\pi_\rel\simeq 0.22\,\mas$, and so $D_L\sim 3.0\,\kpc$. 
In fact, because the fractional errors on both $\theta_\e$ and $\pi_\e$
are relatively large, these estimates will require a more careful
treatment.  However, from the present perspective the main point to note
is that these values 
make the blended light seen in Figure~\ref{fig:cmd} a plausible 
candidate for the lens.

\section{{Blend = Lens?}
\label{sec:blend}}

We therefore begin by gathering the available information about the blend.

\subsection{{Astrometry: Blend is Either The Lens or Its Companion}
\label{sec:astrometry}}

We first measure the astrometric offset between the ``baseline object'' 
and the source, initially finding $\Delta\theta=60\,\mas$ (0.15 pixels),
with the source lying almost due west of the ``baseline object''.
This offset substantially exceeds the formal measurement error 
($\sim 8\,\mas$) based on 
the standard error of the mean of seven near-peak measurements,
as well as our estimate of $\sim 15\,\mas$ for the astrometric
error of the ``baseline object''.   However,
such an offset could easily be induced by 
differential refraction.  That is, the source position is determined
from difference images formed by subtracting the template from 
images near peak, i.e., late in the season when the telescope is always
pointed toward the west, whereas the template is formed from images 
taken over the season (and in any case, the source contributes
less than half the light to these images).  Moreover, the image alignments 
are dominated by foreground main-sequence stars because these are
the brightest in $I$ band.  This contrasts strongly with the situation
for typical
microlensing events for which the majority of bright stars are bulge
giants. Hence, the color offset between
the reference-frame stars and the source is about $\Delta (I-K)\sim 4$.
This means that the mean wavelength of source photons passing through
the $I$-band filter is close to the red edge of this band pass, while
the mean wavelength of reference-frame photons is closer to the middle.
As the effective width of KMT $I$ band is about 160 nm, the
wavelength offset between the two should be about 
$\Delta\lambda\sim 50\,$nm.  Because
blue light has a higher index of refraction than red light, it
appears relatively displaced toward the zenith.  Stated otherwise,
the red light is displaced in the direction of the telescope pointing,
i.e., west.

To quantify this argument, we first review the expected displacement
starting from Snell's Law\footnote{Actually due to Ibn Sahl, circa 984 C.E.}
 ($n=\sin i/\sin r^\prime$), where $n$ is the
index of refraction, $i$ is the angle of incidence, and $r^\prime$ is the
angle of refraction.  We then quantitatively evaluate the
astrometric data within this formalism.  The angular displacement $\delta(i)$
of the source should obey
\begin{equation}
\delta(i) = r^\prime_{\rm source} -r^\prime_{\rm frame}
\simeq {d r^\prime \over d\lambda}\Delta\lambda
\simeq {d \sin r^\prime \over d\lambda}{\Delta\lambda\over \cos i}
\simeq -\tan i {d n\over d\lambda}\Delta\lambda .
\label{eqn:snell}
\end{equation}

\begin{figure}
\plotone{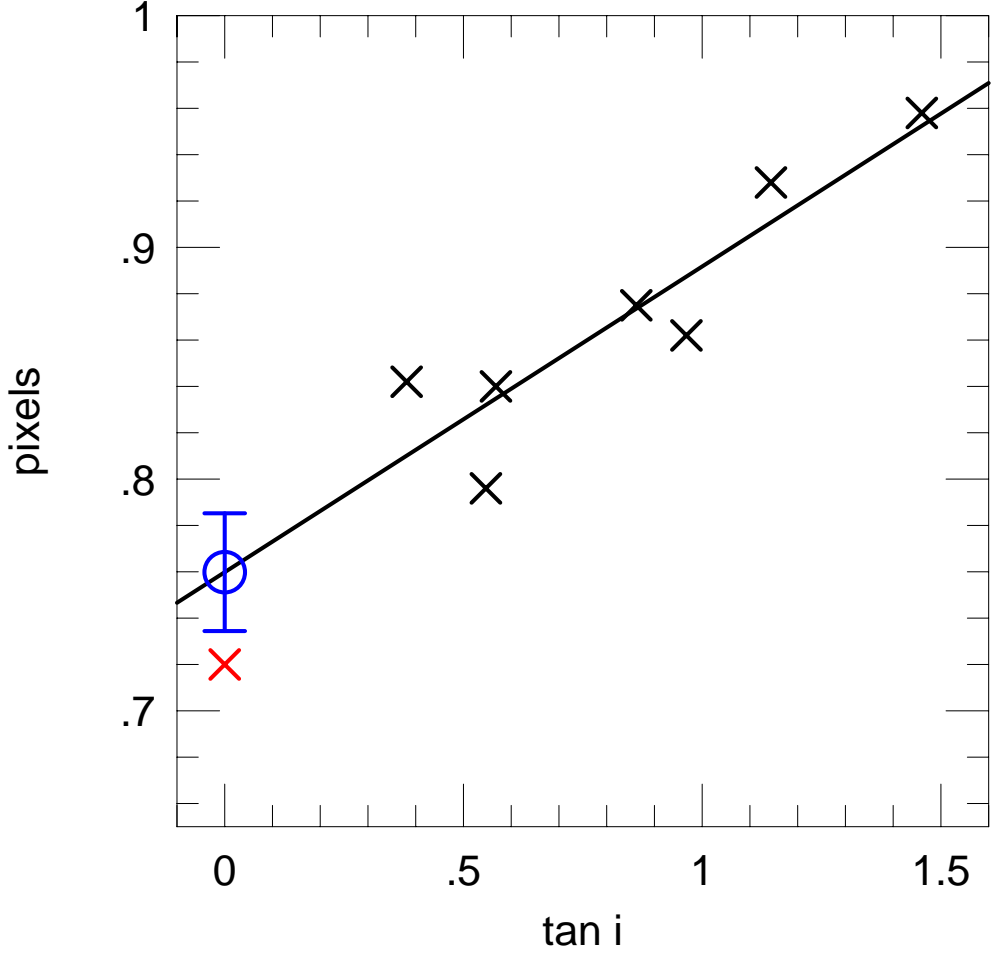}
\caption{Correction for differential refraction along the east-west axis.
Pixel position of the difference-image source in the $x$ (west)
direction as a function of $\tan i$, where $i$ is the angle of incidence
(i.e., airmass = $\sec i$) at seven epochs (black).  The red point
shows the position of the ``baseline object'' on the template.  The
line is a simple regression of the seven points, while the blue circle
is its extrapolation to the zenith.  The agreement within 0.04 pixels
(16 mas), together with similar agreement on $y$ (north-south)
axis, which is not
impacted by differential refraction, shows that the blended light
is either the lens itself or a companion to the lens.  The scatter
of the measurements is 10 mas.  This strong differential fraction is
unusual for the near-standard KMT $I$-band filter and occurs only
because of the extreme reddening, which displaces the mean source light
from the mean reference-frame light within this filter 
by $\Delta\lambda = (41\pm 9)\,$nm.
}
\label{fig:snell}
\end{figure}

Figure~\ref{fig:snell} shows the seven measurements of the $x$
(east-west) coordinate of the source position in pixels versus $\tan i$ in black
and the ``baseline object'' position in red.  The line is a simple
regression without outlier removal.  The scatter about this
line is $\sigma=10\,\mas$ (0.025 pixels).  The $y$ intercept is the
extrapolation of the observed trend to the zenith.  The offset from the
``baseline object''
is only $16\,\mas$ (0.04 pixels), i.e., of order the error in
measuring its position on the template.  The offset in the other (north-south) 
coordinate
(which is not significantly affected by differential refraction) is
likewise $16\,\mas$.  We note that the slope of the line is
$d\theta/d\tan i = (2.56\pm 0.54)\times 10^{-7}\,$radians.
Substituting\footnote{From 
$n-1 = 0.05792105/(238.0185 - (\lambda/\mu{\rm m})^{-2})
+ 0.0016917/(57.362-(\lambda/\mu{\rm m})^{-2})$,
\hfil\break
https://refractiveindex.info/?shelf=other\&book=air\&page=Ciddor\ .}
$dn/d\lambda = -6.17\times 10^{-9}\,{\rm nm}^{-1}$, into
Equation~(\ref{eqn:snell}) yields 
\begin{equation}
\Delta\lambda = \lambda_{\rm source} - \lambda_{\rm frame} =(41\pm 9)\,{\rm nm}.
\label{eqn:deltalambda}
\end{equation}
% (n-1) = 0.05792/(238.0185 - 1/lam^2) + 0.001696/(57.362-1/lam^2)
%dn/dl = 0.05792/(238.0185 - 1/lam^2)^2 + 0.001696/(57.362-1/lam^2)^2
% * 2/lam^3
% (1.0359e-6 + 5.447e-7)*2/0.8**3 = 6.17e-6
%https://refractiveindex.info/?shelf=other&book=air&page=Ciddor

The close proximity of
the baseline object with the source implies that the excess light
is almost certainly associated with the event, i.e., it is either
the lens itself or a companion to the lens or to the source.  That
is, the surface density of stars brighter in $I$ than the blend
is only $90\,{\rm arcmin}^{-2}$.  Hence, the chance of a random alignment
of such a star with the source within $25\,\mas$ is only $\sim 5\times 10^{-5}$.
However, the blend is far too blue to be a companion to the source,
which would require that it be behind the same $E(V-I)\sim 4$ column
of dust.  

\subsection{{Is the Blend a Companion to the Lens?}
\label{sec:dm91}}

Thus, the blend must be either the lens or a companion to the lens.
To evaluate the relative probability of these two options, we should
consider the matter from the standpoint of the blend, which is
definitely in the lens system whether it is the lens or not.
There is a roughly 70\% probability that the blend has a companion,
and if it does, some probability that this companion to the blend
is the lens.  

However, this conditional probability is actually
quite low due to three factors.  We express the arguments in terms
of $Q\ga 1$, the mass ratio of the blend to the host-lens (viewed
as companion to the blend) and $a_b$, the projected separation between
them.  For purposes of this argument, we assume that the lens is
at $D_L\sim 3\,\kpc$, but the final result depends only weakly
on this choice.

First, $a_b<75\,\au$.  Otherwise the astrometric offset between
the source and the ``baseline object'' would be larger than observed.
Second, the source must pass no closer than about 2.5 blend-Einstein-radii
from the blend.  Expressed quantitatively: $a_b> 2.5\,D_L\,\theta_\e Q^{1/2}$.
Smaller separations can be divided into two cases.  Case 1:
$0.5\,D_L\,\theta_\e Q^{1/2}\la a_b < 2.5\,D_L\,\theta_\e Q^{1/2}$. In this
case,the blend would give recognizable microlensing
signatures to the light curve.
Actually, this is a fairly conservative limit because such signatures
will often be present even at larger separations.  Case 2:
$a_b \la 0.5\,D_L\,\theta_\e Q^{1/2}$.  Such cases are possible, but
the planet would then be a circumbinary planet rather than a planet
of the companion to the blend, which would be required to make the blend a 
distinct source of light.  Third, the cross section for lensing
is lower for the blend's putative companion than for the blend itself
by $Q^{-1/2}$.  We take account of all three factors using the binary
statistics of \citet{dm91} and plot the cumulative probability
as a function of host to blend mass ratio in Figure~\ref{fig:dm91}.
The total probability that the blend is a companion to the lens is
only 6.6\%.  

\subsection{{Gaia Proper Motion of the ``Baseline Object''}
\label{sec:gaia}}

Regardless of whether the blend is the lens or a companion to the
lens, the blend proper motion $\bmu_b$ is essentially the same
as that of the lens.  In principle, the two could differ due to orbital
motion.  However, we argued in Section~\ref{sec:dm91} that the projected
separation is at least $a_b\ga 12 Q^{1/2}\,\au$, meaning that the velocity
of the blend relative to the center of mass of the system is less than
$5\,\kms$, which is small compared to the measurement errors in the
problem.

The proper motion of the ``baseline object'' has been measured by
{\it Gaia}
\begin{equation}
\bmu_\base(N,E) = (-3.0,+0.9)\pm (0.8,1.1)\,\masyr,
\label{eqn:gaia}
\end{equation}
with a correlation coefficient of 0.51.  In fact, $\bmu_\base$ is the flux
weighted proper motion of the blend and source in the {\it Gaia} band,
\begin{equation}
\bmu_{\rm base} = (1-\eta)\bmu_B  + \eta\bmu_S 
\rightarrow
(1-\eta)\bmu_L + \eta\bmu_S 
\label{eqn:mubase}
\end{equation}
where $\eta$ is the fraction of total {\it Gaia} flux due to the source.
It may eventually be possible to measure $\eta$ directly from {\it Gaia}
data because there are two somewhat magnified ($A\simeq 1.34$) epochs at
${\rm JD}^\prime = 8342.62$ and 8342.69 and one moderately magnified
($A\simeq 1.75$) epoch at 8364.62.  Based on the reported
photometric error and number of observations, we estimate that
individual {\it Gaia} measurements of the ``baseline object''
have 2\% precision.  If so, {\it Gaia} will determine $\eta$
with fractional precision $\sigma(\eta)/\eta\simeq 0.022/\eta$.
Pending release of {\it Gaia} individual-epoch photometry, we
estimate $\eta$ by first noting that the blend is 0.32 mag 
brighter than the source, even in the $I$ band, and that only the blend
will effectively contribute at shorter wavelengths where the {\it Gaia}
passband peaks.  We therefore estimate that the blend will contribute an equal
number of photons at these shorter wavelengths, while the source will
contribute almost nothing, which implies $\eta=0.27$. 
%1/(2*1.14) =  
%E = (A-1)*eta +- sigma  eta = E/(A-1) ; sigma(eta) = sigma/(A-1)
% sigma_\tot(eta) = sigma/sqrt{sum_i (A_i -1)^2} = 0.02/

We can relate the {\it Gaia} proper motion to the heliocentric proper
motions of the source and lens by writing
\begin{equation}
\bmu_\hel \equiv \bmu_L - \bmu_S;
\qquad
\bmu_\hel = \bmu_\rel + {\pi_\rel\over \au} \bv_{\oplus,\perp},
\label{eqn:bmuhel}
\end{equation}
where $\bv_{\oplus,\perp}(N,E) = (-3.9,-15.0)\,\kms$ is Earth's velocity
projected on the event at $t_0$.  We can then simultaneously solve 
Equations~(\ref{eqn:mubase}) and (\ref{eqn:bmuhel}) to obtain
\begin{equation}
\bmu_L = \eta\bmu_\hel + \bmu_{\rm base};
\qquad
\bmu_S = -(1-\eta)\bmu_\hel + \bmu_{\rm base}.
\label{eqn:bmusol}
\end{equation}

Next, we note that Equation~(\ref{eqn:bmusol}) depends only weakly on
the somewhat uncertain $\pi_\rel$ via the $\bv_{\oplus,\perp}$ term
in Equation~(\ref{eqn:bmuhel}).  For example, if $\pi_\rel = 0.22\,\mas$,
then this term is only $v_{\oplus,\perp}\pi_\rel/\au \sim 0.7\,\masyr$,
which is quite small compared to $\mu_\rel$.  Therefore, to simplify
what follows, we evaluate $\bmu_\hel$ using this value.

\section{{A New Test of the $\bpi_\e$ Measurement}
\label{sec:pietest}}

The {\it Gaia} measurement of the ``baseline object'' and the
resulting Equation~(\ref{eqn:bmusol}) allow us to test the reliability of the
parallax measurement.  Such tests are always valuable, but especially so
in the present case because the modeling of the source variability
could introduce systematic errors into the parallax measurement.  We have
already conducted one test by showing in Table~\ref{tab:evolve} 
that $\bpi_\e$ does
not significantly change as we introduce additional wave-form parameters.
However, the opportunity for additional tests is certainly welcome, particularly
because introducing $\bpi_\e$ only improves the fit by $\Delta\chi^2=13$.

 From a mathematical standpoint, the two degrees of freedom of $\bpi_\e$
can be equally well expressed in Cartesian  $(\pi_{\e,N},\pi_{\e,E})$
or in polar $(\pi_\e,\phi_\pi)$ coordinates.  Here, 
$\tan\phi_\pi\equiv \pi_{\e,E}/\pi_{\e,N}$, i.e., the position angle of
$\bmu_\rel$ north through east.  Cartesian coordinates
are usually more convenient for
light-curve modeling because their covariances are better behaved
(but see \citealt{ob161045}).  However, from a physical standpoint,
polar coordinates are more useful because the amplitude of $\bpi_\e$
contains all the information relevant to $M$ and $\pi_\rel$
(see Equation~(\ref{eqn:massdist})) while the direction contains none.
In particular, a test of the measurement of $\phi_\pi$ that does not
involve any significant assumption about $\pi_\e$ can give added
confidence to the measurement of the latter.

Figure~\ref{fig:path} illustrates such a test.  It shows the 
source and lens proper motions as functions of $\phi_\pi$ in $15^\circ$ steps.  
The cardinal directions are marked in color and labeled.  
The error ellipses (shown for cardinal directions only) take account of both
the {\it Gaia} proper motion error and the uncertainty in the
magnitude of $\mu_\rel$ (at fixed direction).  The cyan ellipses show
the expected dispersions of Galactic-disk (left) and Galactic-bar (right)
sources.  Hence, it is expected that if the parallax solutions are correct,
then at least one of them should yield $\phi_\pi$ that is reasonably 
consistent with one of these two cyan ellipses.  Note that there are
substantial sections of the source ``circle of points'' that would
be inconsistent or only marginally consistent with these ellipses.

The yellow line segments show the ranges of source (outer)
and lens (inner) proper motions implied by the $1\,\sigma$ range of 
the $\phi_\pi$ measurements from the two ($u_0>0$ and $u_0<0$)
solutions.  The source proper motion derived from these solutions
is clearly consistent with a Galactic bar source.  This increases
confidence that $\pi_\e$ is correctly measured within its 
quoted uncertainties as well.

Finally, we note that in order to limit the complexity of 
Figure~\ref{fig:path}, we have fixed both $\pi_\rel=0.22$ and
$\eta=0.27$.  We therefore now consider how this Figure would change
for other values of these quantities.

Changing $\pi_\rel$ by $\Delta\pi_\rel$ would displace the center of 
each ``circle of points'' very slightly, i.e., by 
$-(1-\eta)\Delta\pi_\rel \bv_{\oplus,\perp}\simeq 
(0.06,0.23)(\Delta\pi_\rel/0.1\,\mas)\masyr$ for the source and by
$(-0.02,0.08)(\Delta\pi_\rel/0.1\,\mas)\masyr$ for the lens.
The effect of such a shift on this figure would hardly be discernible.

Changing $\eta$, for example from 0.27 to 0.22 or 0.32, would make the
source ``circle of points'' larger or smaller by 7\%.  Again, such
changes would hardly impact the argument given above.

\section{{Physical Parameters}
\label{sec:phys}}

While both $\theta_\e$ and $\pi_\e$ are measured, they have relatively
large fractional errors: of order 20\% and 25\%, respectively.  Hence, it is
inappropriate to evaluate the physical parameters simply by algebraically
propagating errors, using for example, Equation~(\ref{eqn:massdist}).
Instead, we evaluate all physical quantities by applying these (and
other) algebraic equations to the output of the MCMC.  The results
are tabulated in Table~\ref{tab:phys} and illustrated in Figure~\ref{fig:phys}.
Because the source proper motion is consistent with Galactic-bar
(but not Galactic-disk) kinematics, we simply assign the source
distance $D_S=9\,\kpc$.  See Section~\ref{sec:pietest} and 
Figure~\ref{fig:path}.  The errors are relatively large, but based on the
microlensing data alone, the lens is likely to be an F or G star, with
a super-Jovian planet.  

This result is supported by the fact that
the blend (lens) lies near the ``bottom edge''
(alternatively ``blue edge'') of the foreground main-sequences stars
on the CMD (Figure~\ref{fig:cmd}).  To understand the implications of this
position, consider two stars of the same apparent color $(V-I)$, 
but which differ
in reddening by $\Delta E(V-I)$ and in intrinsic color by $\Delta(V-I)_0$.
Tautologically, $\Delta E(V-I) + \Delta(V-I)_0 = 0$.  We then adopt
estimates $\Delta A_I = 1.25\Delta E(V-I)$ and $\Delta M_I = 2.3\Delta(V-I)$.
This leads to an estimate
\begin{equation}
\Delta I = \Delta M_I + \Delta A_I + \Delta{\rm DM}
= -0.84\Delta A_I + \Delta{\rm DM},
\label{eqn:dmod}
\end{equation}
where $\Delta{\rm DM}$ is the difference in distance modulus.

Now, $A_I$ is roughly linear in distance $A_I = 5.2\,{\rm mag}/(9\,\kpc) 
= 0.58\,{\rm mag\,kpc^{-1}}$, while DM is logarithmic, 
$d{\rm DM}/d D= (5/\ln 10)\,D^{-1}$.  Hence, the derivatives of the two
terms in Equation~(\ref{eqn:dmod}) are equal and opposite at 
$D_{\rm stationary}\simeq 4.45\,\kpc$.  As the second derivative
of Equation~(\ref{eqn:dmod}) is strictly negative, this stationary
point is a maximum.  That is, the bottom of the foreground track in the
CMD corresponds roughly to stars at this distance, which implies 
that the lens/blend has 
$D_L\sim D_{\rm stationary}$, $A_{I,L}\sim 2.6$, and $M_{I,L}\sim 2.9$.  
This would be consistent with an $M\sim 1.5\,M_\odot$ main-sequence
star, or perhaps a star of somewhat lower mass on the turn off (which
is not captured by the simplified formalism of Equation~(\ref{eqn:dmod})).
That is, this qualitative argument is broadly consistent with the
results in Table~\ref{tab:phys}.
We discuss how followup observations can
improve the precision of these estimates in Section~\ref{sec:followup}.

%[b - (zsun/r0 + bsgr)]*Dl + zsun
%[b-bsgr]*DL + zsun*(1 - DL/R0)
We note that at the distances indicated in Figure~\ref{fig:phys} (or
by this more qualitative argument), the lens lies quite close to
the Galactic plane,
\begin{equation}
z_L = z_\odot\biggl(1-{D_L\over R_0}\biggr) + 
D_L\sin(b-b_{\rm sgrA*}) = -0.0060(D_L - 2.48\,\kpc),
\label{eqn:zperp}
\end{equation}
where $b_{\rm sgrA*}$ is the Galactic latitude of SgrA*, $R_0$ is the
Galactocentric distance, and where we 
have adopted $z_\odot = 15\,\pc$ for the height of the Sun above
the Galactic plane.  That is, if $D_L$ is within a kpc of 2.48 kpc,
then the lens is within 6 pc, of the Galactic plane.

\section{{Discussion}
\label{sec:discuss}}

\subsection{{Lowest Galactic-Latitude Planet}
\label{sec:lowb}}

At $b=-0.28$, KMT-2018-BLG-1292Lb is the lowest Galactic-latitude 
microlensing planet yet detected.  Yet, KMTNet did not consciously
set out to monitor the Galactic plane.  Instead, it has a few fields,
including BLG13, BLG14, BLG18, BLG38, and BLG02/BLG42, whose
corners ``inadvertently'' cross the Galactic plane or come very close
to it.  See Figure~\ref{fig:fields}.
This is a side effect of having a large-format square
camera on an equatorial mount telescope (together with the fact that
the Galactic plane is inclined by $\sim 30^\circ$ relative to north
toward the Galactic center).  Of these five fields, BLG13 has
the lowest cadence ($\Gamma = 0.15$--$0.2\,{\rm hr}^{-1}$), with
BLG14 and BLG18 being 5 times higher and BLG02/42 being 20 times higher.
Nevertheless, despite this low-cadence (further aggravated by
the fact that the anomaly occurred near the end of the season,
when the Galactic bulge was visible for only a few hours per night)
and the very high extinction $A_I\sim 5.2$, KMT-2018-BLG-1292Lb
is reasonably well characterized, with measurements of both $\theta_\e$
and $\bpi_\e$.  This leads us to assess the reason for this
serendipitous success.

The first point is that the source is very luminous and very red,
which together made the event reasonably bright in spite of the
high extinction.  It also implies a large source radius, with
a source-diameter crossing time of almost one day, $2 t_* = 19\,$hr.
Hence, despite the low effective combined cadence from all three
observatories $\Gamma\sim 1\,{\rm day}^{-1}$, the source profiles
on the source plane nearly overlap as it transits the caustic.
See Figure~\ref{fig:caustic}.  Thus, although the actual trajectory
fortuitously rides the edge of a caustic, even random trajectories
through the caustic would have led to significant finite source
effects for some measurements, and therefore to a measurement of
$\theta_\e$.  This large source size is not fortuitous: in high
extinction fields, such large sources are the only ones that
will give rise to detectable microlensing events in the optical, 
apart from a handful of very high magnification events.  That is, although
high-extinction fields necessarily greatly reduce the number
of sources that can be probed for microlensing events, those
that can shine through the dust can yield well-characterized
events even with very low cadence.  This means that optical surveys
could in principle more systematically probe the Galactic plane
for microlens planets at relatively low cost in observing
time.

Although Figure~\ref{fig:fields} is presented primarily to show
current optical coverage of the Galactic plane and to illustrate
the possibilities for future coverage, it also has more general implications
for understanding past and possible future strategies for microlensing
planet detection. We summarize these here.  The colored circles
in Figure~\ref{fig:fields} 
represent published microlensing planets discovered in
2003-2017, while the black squares show 2018 event locations that we
assess as likely to yield future planet publications.  The blue
points, which are from 2003-2010, i.e.,  prior to OGLE-IV, are 
uniformly distributed over the southern bulge.  By contrast (and
restricting attention for the moment to the southern bulge) planet
detections in all subsequent epochs are far more concentrated toward
the regions near $(l,b)\sim (+1,-2.5)$.  During 2003-2010, the cadence
of the survey observations was typically too low to detect and 
characterize planets by themselves\footnote{However, note that even
in this period, six of the 22 planetary events were detected and 
characterized in pure survey mode: MOA-2007-BLG-192, MOA-bin-1,
MOA-2008-BLG-379, OGLE-2008-BLG-092, OGLE-2008-BLG-355, MOA-2010-BLG-353
\citep{mb07192,moabin1,mb08379,ob08092,ob08355,mb10353}.}.
Hence, most planets were discovered
by a combination of follow-up observations (including survey auto-follow-up)
and survey observations of events alerted by 
OGLE and/or MOA.  The choice of these follow-up efforts was not
strongly impacted by survey cadence, which in any case was relatively
uniform.  It is still slightly surprising that the planet detections
do not more closely track the underlying event rate, which is higher
toward the concentration center of later planet detections.

As soon as the OGLE-IV survey started (green points 2011-2013),
the overall detection rate increases by a factor 2.7, but
the southern-bulge planets also immediately become more concentrated.
This partly reflects that the OGLE and MOA surveys (together with
the Wise survey, \citealt{shvartzvald16}) 
were very capable of detecting planets without followup
observations in their higher-cadence regions, which were near this
concentration.  But in addition, these higher-cadence regions 
began yielding vastly more alerted events and also better characterization
of these events, which also tended to concentrate the targets for
follow-up observations.  Also notable in this period are the first
three planets in the northern bulge, to which OGLE-IV devoted a few
relatively high-cadence fields.

In the next period (yellow points, 2014-2015), the surveys remained
similar, but follow-up observations were sharply curtailed due
to reduction of work by the Microlensing Follow-Up Network
($\mu$FUN, \citealt{gould10}).  The rate drops by 45\%,
but the main points to note
are that the southern bulge discoveries become even more concentrated
and there are no northern bulge discoveries.  In particular, comparing
2003-2010 with 2011-2015, the dispersion in the $l$ direction 
in the southern bulge drops by more than a factor two, from
$3.21^\circ\pm 0.50^\circ$ to $1.45^\circ\pm 0.20^\circ$.

The magenta and black points together show the planets discovered
during the three years when the KMT wide-area survey joined the
ongoing OGLE and MOA surveys, which is also
the first time that the KMT fields shown in the figure become relevant
to the immediate discussion.
There are several points to note.  First, the rate of detection
increases by a factor 2.7 relative to the previous two years 
(or by a factor 1.8 relative to the previous five years).  Second, the
southern bulge planets become somewhat less concentrated, but
still tend to follow the KMT very-high-cadence (numbered in red)
and high-cadence (numbered in magenta) fields.  In fact, only four out
of 24 planets in the southern bulge lie outside of these fields.
This should be compared to the 22 blue (2003-2010) points, 11 (half) of
which lie outside these fields.  Finally, there are  eight planets
in the northern bulge, all in the four high cadence fields.

This history seems to indicate that there is substantial potential for
finding microlensing planets in low cadence fields by carrying out
aggressive follow-up observations similar to those of the pre-OGLE-IV era.

\subsection{{Precise Lens Characterization From Spectroscopic Followup}
\label{sec:followup}}

\begin{figure}
\plotone{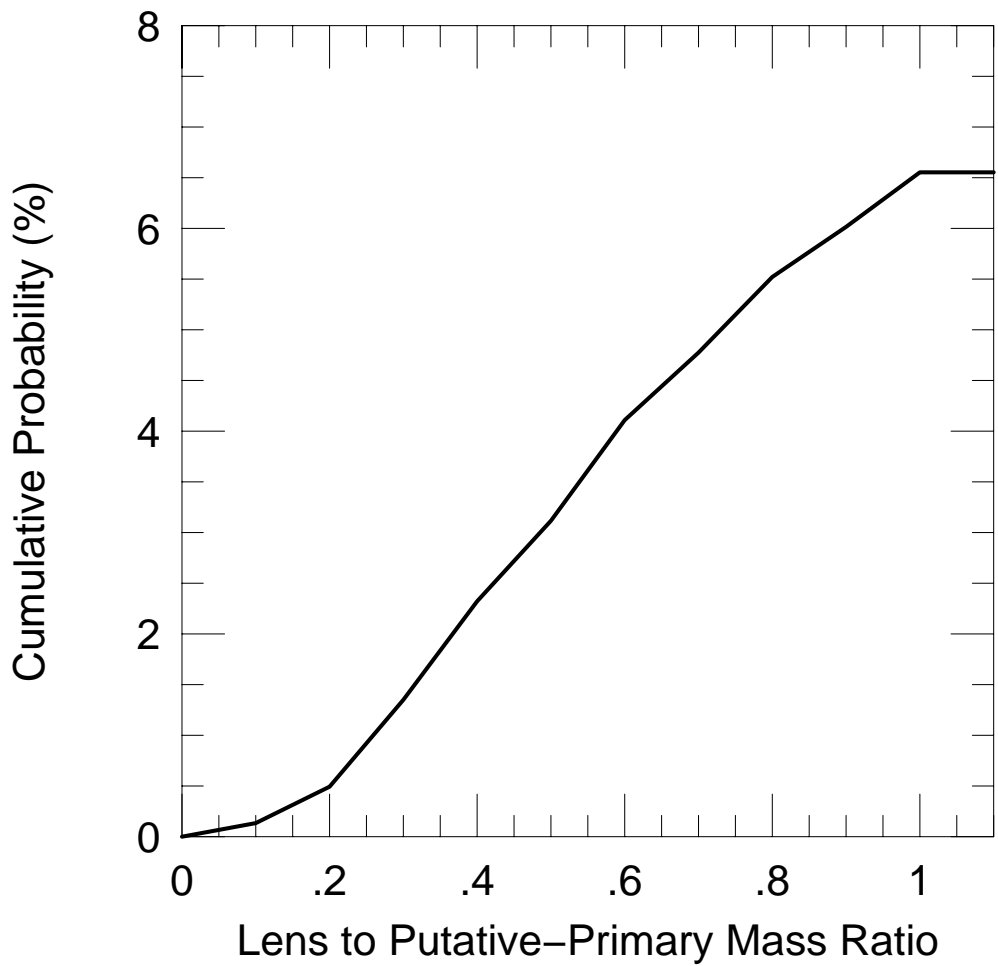}
\caption{Cumulative probability that the host is a companion to
the blend (rather than the blend itself) as function of lens to
``putative primary'' (blend) mass ratio.  Although lower mass
secondaries of G-dwarf binaries are more common \citep{dm91},
these are suppressed by lower cross sections ($\propto M^{1/2}$)
and smaller range of semi-major axis in which the ``putative primary''
could avoid giving rise to microlensing signatures.  The total probability
that the lens is a companion to the blend (rather than the blend itself)
is only 6.6\%.
}
\label{fig:dm91}
\end{figure}

\begin{figure}
\plotone{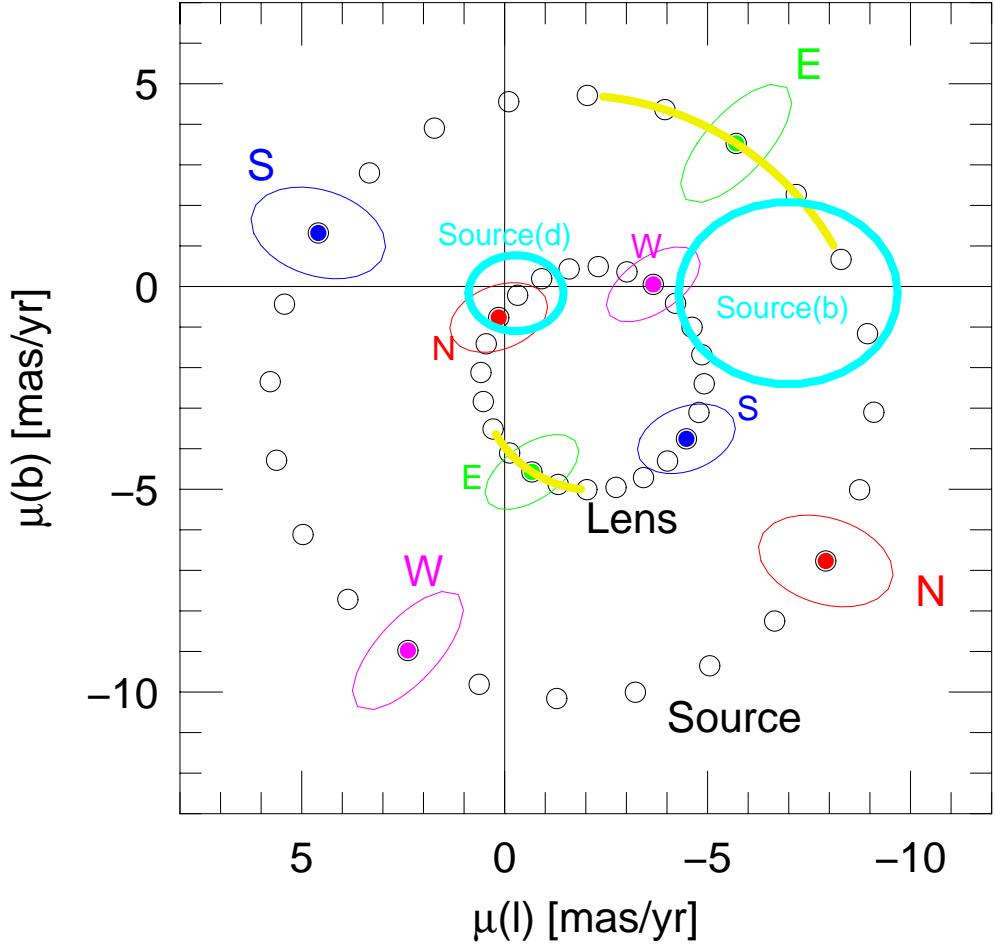}
\caption{
Proper motion of the lens and source, under the assumption
that the angle of the parallax vector $\bpi_\e$ has a direction
indicated by the figure labels, north (red), east (green), south (blue),
and west (magenta), with $15^\circ$ steps indicated by black circles.
The error ellipses, which take account of both the {\it Gaia} errors
and correlation coefficient and the error in the magnitude of the
geocentric lens-source relative proper motion, $\mu_\rel$, are shown
for the cardinal directions.  The cyan ellipses show the expected
proper motion dispersions for disk (left) and bar (right) sources.
The $1\,\sigma$ range of the measured source proper 
motion (upper yellow track), which is derived from the direction $\phi_\pi$ 
of the microlens parallax $\bpi_\e$, is consistent with the
kinematics of the Galactic-bar.  This lends support to the other
polar coordinate of the parallax vector, i.e., its amplitude $\pi_\e$,
being correctly measured as well.
}
\label{fig:path}
\end{figure}

\begin{figure}
\plotone{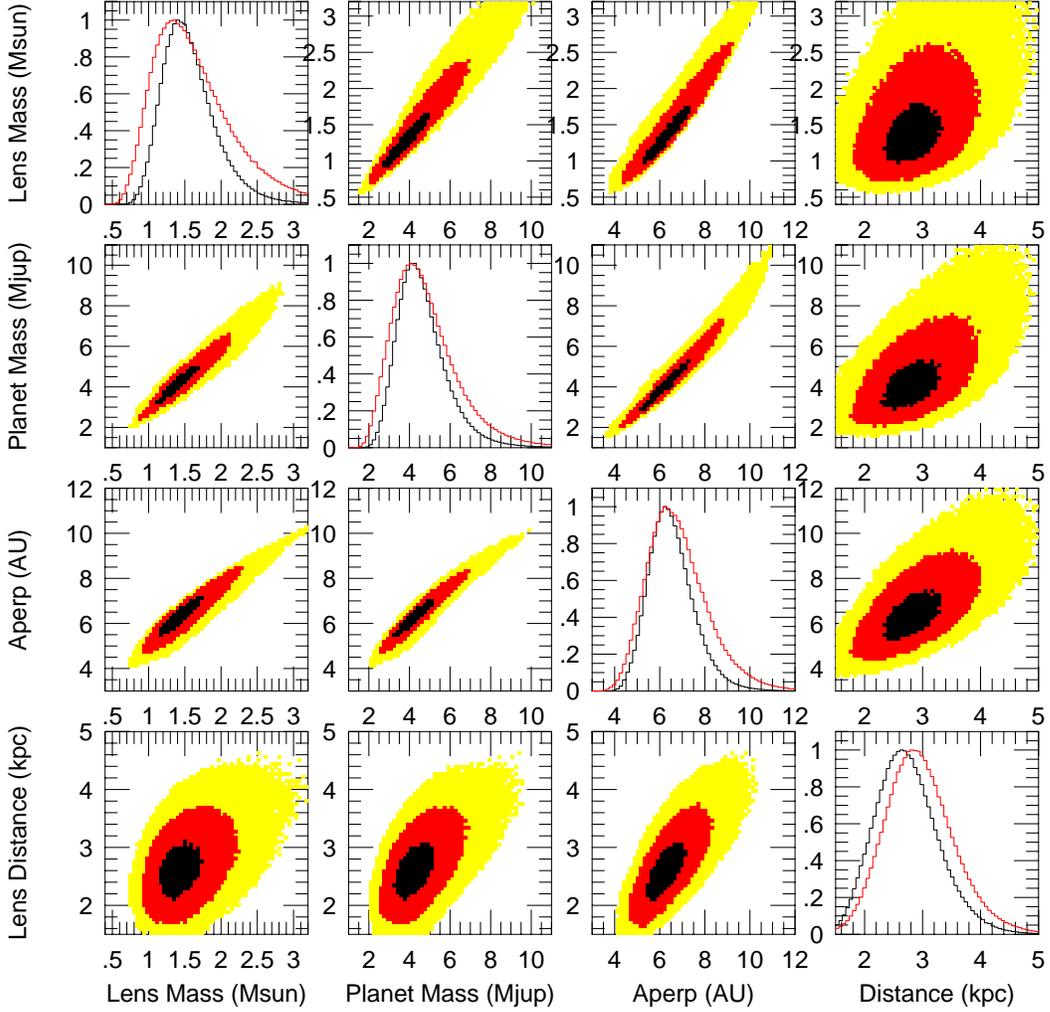}
\caption{Likelihood distributions for pairs of physical parameters,
$(M,M_p,a_\perp,D_L)$, i.e., the lens mass, the planet mass, the
host-planet projected separation, and the distance to the lens system.
The lower-left panels show the $(u_0<0)$ solution, while
the upper-right panels show the $(u_0>0)$ solution.  Black, red, and
yellow show likelihood ratios 
$[-2\Delta\ln({\cal L}/{\cal L}_{\rm max})]<(1,4,9)$, respectively.
The diagonal shows the single-parameter histograms, with 
$(u_0<0)$ in black and $(u_0>0)$ in red.
}
\label{fig:phys}
\end{figure}

\begin{figure}
\plotone{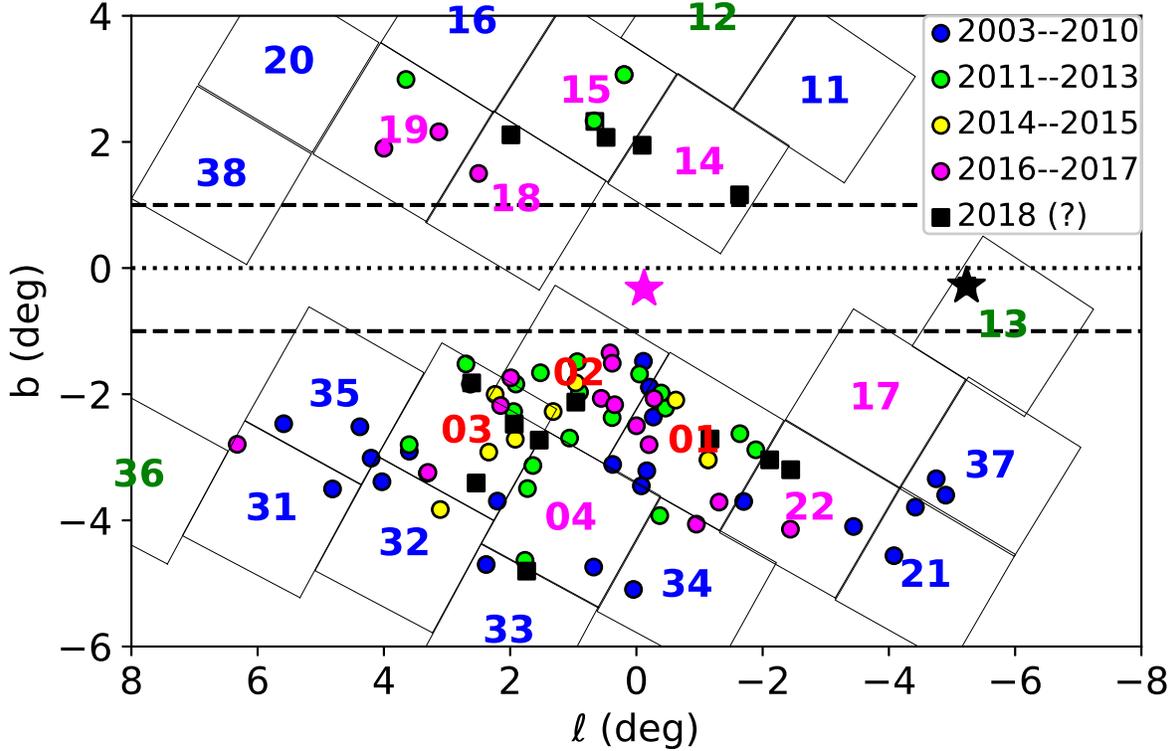}
\caption{
The positions of published microlensing planets from 2003-2017 (circles)
and likely-to-be-published microlensing planets from 2018 (squares) are
shown against the KMT field configuration for 2016-2018,
which are color-coded (red, magenta, blue, green) according to
their nominal cadence $\Gamma = (4,1,0.4,0.2)\,{\rm hr}^{-1}$.  The two
planets that lie close to the plane, UKIRT-2017-BLG-001Lb \citep{ub17001}
and KMT-2018-BLG-1292Lb (this work) are shown as stars. Despite the fact
that KMT systematically avoids the Galactic plane, five of its fields
(BLG13, 14, 18, 38, and 02/42) ``inadvertently'' cross the plane or come 
close to it.  The detection of KMT-2018-BLG-1292Lb in the lowest cadence
of these fields suggests that the Galactic plane could be probed for
planets in the optical at relatively low cost.  Published planets are
color coded by year of discovery.  Their changing areal distribution with
time is discussed in Section~\ref{sec:lowb}.
}
\label{fig:fields}
\end{figure}

As shown in Section~\ref{sec:astrometry}, the blend is almost certainly either
the lens or its companion and as shown in Section~\ref{sec:dm91}, it is very
likely to be the lens.  See Figure~\ref{fig:dm91}.  Hence, a 
medium-resolution spectrum of the blend would greatly clarify the nature
of the lens in two ways.  

First, by spectrally typing the blend one could
obtain a much better estimate of its mass.  Second, if the mass turns out
to be, e.g., $M\sim 1.5\,M_\odot$ in line with the results in 
Table~\ref{tab:phys}, then this would further reduce the
probability that the lens is a companion to the blend relative to the
6.6\% probability that we derived in Section~\ref{sec:dm91}.  This
is because companions to the blend with mass ratio $Q^{-1}\la 0.5$
would then have masses $M\la 0.75\,M_\odot$, which are significantly disfavored
by the results of Section~\ref{sec:phys}.  Hence, of order half the
probability allowed by Figure~\ref{fig:dm91} would be eliminated, which
would further increase confidence that the blend (now spectrally typed)
was the lens.

Such a spectrum could be taken immediately.  Of course, the source
would remain in the aperture for many years, but it is unlikely to
contribute much light in the $V$- and $R$-band ranges of the spectrum,
as we discussed in Section~\ref{sec:astrometry}.  In addition, the source
spectrum is likely to be displaced by many tens of $\kms$ from that of
the blend.

%\begin{equation}
%\label{eqn:}
%\end{equation}

\acknowledgments 
We thanks Christopher Kochanek for providing SMARTS ANDICAM $I/H$ data.
AG was supported by AST-1516842 from the US NSF
and by JPL grant 1500811.
Work by CH was supported by grant 2017R1A4A1015178 of the National Research Foundation of Korea.
This research has made use of the KMTNet system operated by the Korea
Astronomy and Space Science Institute (KASI) and the data were obtained at
three host sites of CTIO in Chile, SAAO in South Africa, and SSO in
Australia.
We gratefully acknowledge the use of data from the ESO Public Survey program IDs 179.B-2002 and 198.B-2004 taken with the VISTA telescope, and data products from the Cambridge Astronomical Survey Unit (CASU).
D.M. gratefully acknowledges support provided by the Ministry for the Economy, Development and Tourism, Programa Iniciativa Cientifica Milenio grant IC120009, awarded to the Millennium Institute of Astrophysics (MAS), by the BASAL Center for Astrophysics and Associated Technologies (CATA) through grant AFB-170002, and by project Fondecyt No. 1170121.
R.K.S. acknowledges support from CNPq/Brazil through
through projects 308968/2016-6 and 421687/2016-9.
J.A-G. acknowledges support by the Ministry of Economy, Development, and Tourism's Millennium Science Initiative through grant IC120009, awarded to the Millennium Institute of Astrophysics (MAS).

\begin{deluxetable}{lcccc}
\tablecolumns{5} \tablewidth{0pc} \tablecaption{\textsc{Best Fit
Models of KMT-2018-BLG-1292}} \tablehead{ \colhead{} &
\colhead{} & \colhead{} & \multicolumn{2}{c}{Parallax $(P_{2})$} \\
\cline{4-5} \colhead{Parameters} & \colhead{Standard} &
\colhead{Standard $(P_{2})$} & \colhead{$u_0>0$} & \colhead{$u_0<0$}
} \startdata
  $\chi^2/\rm{dof}$              &750.637/721          &723.357/715          &710.145/713           &710.294/713          \\
  $t_0$ $(\rm{HJD}^{\prime})$    &8408.91 $\pm$ 0.52 &8408.98 $\pm$ 0.52 &8408.35 $\pm$ 0.15  &8407.91 $\pm$ 0.33 \\
  $u_0$                          &0.268 $\pm$ 0.009    &0.269 $\pm$ 0.009    &0.286 $\pm$ 0.013     &-0.281 $\pm$ 0.008   \\
  $t_{\rm E}$ $(\rm{days})$      &67.33 $\pm$ 1.57   &66.48 $\pm$ 1.50   &61.87 $\pm$ 2.05    &60.80 $\pm$ 1.36   \\
  $s$                            &1.328 $\pm$ 0.009    &1.333 $\pm$ 0.009    &1.347 $\pm$ 0.013     &1.343 $\pm$ 0.008    \\
  $q$ $(10^{-3})$                &2.671 $\pm$ 0.245    &2.705 $\pm$ 0.245    &2.852 $\pm$ 0.270     &2.982 $\pm$ 0.221    \\
  $\alpha$ $(\rm{rad})$          &2.595 $\pm$ 0.009    &2.601 $\pm$ 0.009    &2.576 $\pm$ 0.020     &-2.589 $\pm$ 0.020   \\
  $\rho$ $(10^{-3})$             &5.790 $\pm$ 0.821    &5.687 $\pm$ 0.776    &6.505 $\pm$ 1.135     &6.516 $\pm$ 0.777    \\
  $\pi_{\rm{E},\it{N}}$          &-                    &-                    &0.032 $\pm$ 0.058     &-0.021 $\pm$ 0.061   \\
  $\pi_{\rm{E},\it{E}}$          &-                    &-                    &0.118 $\pm$ 0.029     &0.131 $\pm$ 0.027    \\
  $f_S$                          &0.359 $\pm$ 0.015    &0.365 $\pm$ 0.015    &0.387 $\pm$ 0.021     &0.379 $\pm$ 0.013    \\
  $f_B$                          &0.450 $\pm$ 0.015    &0.446 $\pm$ 0.014    &0.421 $\pm$ 0.021     &0.430 $\pm$ 0.014    \\
  $t_*$ $(\rm{days})$            &0.390 $\pm$ 0.049    &0.378 $\pm$ 0.046    &0.403 $\pm$ 0.061     &0.396 $\pm$ 0.044    \\
  $a_1$                          &                     &0.012 $\pm$ 0.004    &0.008 $\pm$ 0.003     &0.010 $\pm$ 0.003    \\
  $P_1$ $(\rm{days})$            &-                    &70.13 $\pm$ 9.81   &62.24 $\pm$ 9.71   &63.63 $\pm$ 9.22  \\
  $\phi_1$                       &-                    &0.670 $\pm$ 0.513    &0.993 $\pm$ 0.611     &0.451 $\pm$ 0.577    \\
  $a_2$                          &-                    &0.009 $\pm$ 0.003    &0.007 $\pm$ 0.002     &0.007 $\pm$ 0.002    \\
  $P_2$ $(\rm{days})$            &-                    &13.04 $\pm$ 2.32   &13.00 $\pm$ 4.90   &13.00 $\pm$ 6.85  \\
  $\phi_2$                       &-                    &-0.070 $\pm$ 0.324   &-0.450 $\pm$ 0.549    &0.076 $\pm$ 0.773    \\
\enddata
%\tablecomments{} 
\label{tab:ulens}
\end{deluxetable}

\begin{deluxetable}{lccc}
\tablecolumns{4} \tablewidth{0pc} \tablecaption{\textsc{Parameter
Evolution with Additional Periodic Components}} \tablehead{
\colhead{Parameters} & \colhead{$P_{0}$} & \colhead{$P_{1}$} &
\colhead{$P_{2}$} } \startdata
  $\rm{Parallax} (u_0>0)$ &&&\\
  $\chi^2/\rm{dof}$              &731.918/719          &719.868/716          &710.145/713           \\
  $q$ $(10^{-3})$                &2.781 $\pm$ 0.240    &2.865 $\pm$ 0.223    &2.852 $\pm$ 0.270     \\
  $\rho$ $(10^{-3})$             &5.809 $\pm$ 0.871    &5.433 $\pm$ 0.831    &6.505 $\pm$ 1.135     \\
  $f_S$                          &0.363 $\pm$ 0.015    &0.368 $\pm$ 0.015    &0.387 $\pm$ 0.021     \\
  $\pi_{\rm{E},\it{N}}$          &0.0002 $\pm$ 0.053   &0.088 $\pm$ 0.058    &0.032 $\pm$ 0.058     \\
  $\pi_{\rm{E},\it{E}}$          &0.105 $\pm$ 0.027    &0.119 $\pm$ 0.028    &0.118 $\pm$ 0.029     \\
  $a_1$                          &-                    &0.009 $\pm$ 0.003    &0.008 $\pm$ 0.003     \\
  $P_1$ $(\rm{days})$            &-                    &62.17 $\pm$ 9.26  &62.24 $\pm$ 9.71   \\
  $\phi_1$                       &-                    &0.705 $\pm$ 1.234    &0.993 $\pm$ 0.611     \\
  $a_2$                          &-                    &-                    &0.007 $\pm$ 0.002     \\
  $P_2$ $(\rm{days})$            &-                    &-                    &13.00 $\pm$ 4.90   \\
  $\phi_2$                       &-                    &-                    &-0.450 $\pm$ 0.549    \\
  \cline{1-4}
  $\rm{Parallax} (u_0<0)$ &&&\\
  $\chi^2/\rm{dof}$              &732.186/719          &719.252/716          &710.294/713          \\
  $q$ $(10^{-3})$                &2.937 $\pm$ 0.247    &3.117 $\pm$ 0.236    &2.982 $\pm$ 0.221    \\
  $\rho$ $(10^{-3})$             &6.083 $\pm$ 0.930    &6.728 $\pm$ 0.909    &6.516 $\pm$ 0.777    \\
  $f_S$                          &0.368 $\pm$ 0.016    &0.381 $\pm$ 0.016    &0.379 $\pm$ 0.013    \\
  $\pi_{\rm{E},\it{N}}$          &0.003 $\pm$ 0.054    &-0.008 $\pm$ 0.058   &-0.021 $\pm$ 0.061   \\
  $\pi_{\rm{E},\it{E}}$          &0.118 $\pm$ 0.028    &0.129 $\pm$ 0.027    &0.131 $\pm$ 0.027    \\
  $a_1$                          &-                    &0.008 $\pm$ 0.003    &0.010 $\pm$ 0.003    \\
  $P_1$ $(\rm{days})$            &-                    &62.67 $\pm$ 13.16  &63.63 $\pm$ 9.22  \\
  $\phi_1$                       &-                    &0.453 $\pm$ 1.346    &0.451 $\pm$ 0.577    \\
  $a_2$                          &-                    &-                    &0.007 $\pm$ 0.002    \\
  $P_2$ $(\rm{days})$            &-                    &-                    &13.00 $\pm$ 6.85  \\
  $\phi_2$                       &-                    &-                    &0.076 $\pm$ 0.773    \\
  \enddata
%\tablecomments{} 
\label{tab:evolve}
\end{deluxetable}

\begin{deluxetable}{lccccc}
\tablecolumns{6} \tablewidth{0pc} \tablecaption{\textsc{Physical
parameters}} \tablehead{ \colhead{} & \multicolumn{2}{c}{Parallax
$(P_{2})$} &
\\
\cline{2-3} \colhead{Quantity} & \colhead{$u_0>0$} &
\colhead{$u_0<0$} } \startdata
  $M_{\rm lens}$ $[M_\sun]$       &$1.54_{-0.43}^{+0.67}$       &$1.51_{-0.30}^{+0.41}$     \\
  $M_{\rm planet}$ $[M_J]$        &$4.53_{-1.26}^{+1.79}$       &$4.45_{-0.98}^{+1.32}$     \\
  $a_{\bot}$ [au]               &$6.65_{-1.14}^{+1.47}$       &$6.41_{-0.88}^{+1.07}$     \\
  ${\it D_L}$ [kpc]            &$2.92_{-0.54}^{+0.62}$       &$2.69_{-0.51}^{+0.58}$     \\
  $\theta_{\rm{E}}$ [mas]        &$1.71_{-0.26}^{+0.34}$       &$1.80_{-0.23}^{+0.28}$     \\
  $\mu_{{\rm hel},N}$ [mas/yr]    &$3.6_{-4.5}^{+2.9}$       &$-5.2_{-3.5}^{+4.8}$    \\
  $\mu_{{\rm hel},E}$ [mas/yr]    &$8.1_{-1.8}^{+1.8}$       &$8.2_{-2.3}^{+1.7}$     \\
  $v_{{\rm L,LSR},l}$ [km/s]       &$9.5_{-12.5}^{+6.6}$      &$-13.0_{-9.8}^{+14.0}$  \\
  $v_{{\rm L,LSR},b}$ [km/s]     &$-44.7_{-17.3}^{+17.0}$   &$-52.4_{-11.0}^{+10.2}$ \\
 \enddata
%\tablecomments{} 
\label{tab:phys}
\end{deluxetable}

\begin{thebibliography}{99}

%\bibitem[Afonso et al.(2001)]{eb20005} Afonso, C., Albert, J.N., Anderson, J. et al., 2001, \aap, 378, 1014

\bibitem[Alard \& Lupton(1998)]{alard98} Alard, C. \& Lupton, R.H.,1998, \apj, 503, 325

%\bibitem[Albrow et al.(2000)]{mb9741}Albrow, M.\ Beaulieu, J.-P., Caldwell, J.A.R.,, et al.\ 2000, \apj, 534, 894

%\bibitem[Albrow et al.(2001)]{albrow01}Albrow, M.D., An, J., Beaulieu, J.-P., et al.\ 2000, \apjl, 556, 113

\bibitem[Albrow et al.(2009)]{albrow09}Albrow, M.\ D., Horne, K., Bramich, D.\ M., et al.\ 2009, \mnras, 397, 2099


\bibitem[An \& Gould(2001)]{angould}An, J.H., \& Gould, A. 2001, \apj, 563, L111

%\bibitem[Bachelet et al. (2012)]{bachelet12}Bachelet, E., Fouqu{\'e}, P., Han, C., et al.\ 2012, \aap, 547, 55


\bibitem[Batista et al.(2011)]{mb09387} Batista, V., Gould, A., Dieters, S. et al. \aap, 529, 102

%\bibitem[Batista et al.(2014)]{mb11293b} Batista, V., Beaulieu, J.-P., Gould, A., et al. \apj, 780, 54

%\bibitem[Batista et al.(2015)]{ob05169bat} Batista, V., Beaulieu, J.-P., Bennett, D.P., et al. 2015, \apj, 808, 170

%\bibitem[Beaulieu et al.(2006)] {ob05390}Beaulieu, J.-P. Bennett, D.P., Fouqu\'e, P. et al. 2006, Nature, 439, 437

%\bibitem[Bennett \& Rhie(1996)]{bennett96}  Bennett, D.P. \& Rhie, S.-H. 1996, \apj, 472, 660

\bibitem[Bennett et al.(2008)]{mb07192}  Bennett, D.P., Bond, I.A., Udalski, A., et al.\ 2008, \apj, 684, 663

%\bibitem[Bennett et al.(1999)]{mb9741x}  Bennett, D.P., Rhie, D.P., Becker, A.C., et al.\ 1999, Natur, 402 57

\bibitem[Bennett et al.(2012)]{moabin1}  Bennett, D.P., Sumi, T., Bond, I.A., et al.\ 2012, \apj, 757, 119

%\bibitem[Bennett et al.(2015)]{ob05169ben} Bennett, D.P., Bhattacharya, A., Anderson, J., et al. 2015, \apj, 808, 169

\bibitem[Bensby et al.(2013)]{bensby13} Bensby, T. Yee, J.C., Feltzing, S.\ et al.\ 2013, \aap, 549A, 147

\bibitem[Bessell \& Brett(1988)]{bb88} Bessell, M.S., \& Brett, J.M.\ 1988, \pasp, 100, 1134

%\bibitem[Bond et al.(2004)]{ob03235} Bond, I.A., Udalski, A., Jaroszy\'nski, M. et al.\ 2004, \apj, 606, L155

%\bibitem[Bond et al.(2017)]{ob161195a} Bond, I.A., Bennett, D.P., Sumi, T. et al.\ 2017, \mnras, 469, 2434

%\bibitem[Bozza et al.(2012)]{bozza12}Bozza, V., Dominik, M., Rattenbury, N.\ J., et al.\ 2012, \mnras, 424, 902

%\bibitem[Bramich(2008)]{bramich08} Bramich, D. M. 2008, \mnras, 386, L77.

%\bibitem[Calchi Novati et al.(2015)]{21event}  Calchi Novati, S., Gould, A., Udalski, A., et al., 2015, \apj, 804, 20

%\bibitem[Calchi Novati et al.(2015)]{170event} Calchi Novati, S., Gould, A., Yee, J.C., et al. 2015, \apj, 814, 92

%\bibitem[Cassan et al.(2012)]{cassan12} Cassan, A., Kubas, D., Beaulieu, J.-P., et al., 2012, Nature, 481, 167

%\bibitem[Chang-Refsdal(1979)]{cr1}Chang, K.\ \& Refsdal, S.\ 1979, Nature, 282, 561

%\bibitem[Chang \& Refsdal(1984)]{cr2}Chang, K.\ \& Refsdal, S.\ 1984, \aap, 130, 157

%\bibitem[Choi et al.(2013)]{choi13}Choi, J.-Y., Han, C., Udalski, A., et al.\ 2013, \apj, 768, 129

%\bibitem[Chung et al.(2017)]{ob151482} Chung, S.-J., Zhu, W., Udalski, A., 2017, \apj, 838, 154

%\bibitem[Clanton \& Gaudi(2014a)]{clanton14}Clanton, C.D. \& Gaudi, B.S. 2014a, \apj, 791, 90

%\bibitem[Clanton \& Gaudi(2014b)]{clanton14b}Clanton, C.D. \& Gaudi, B.S. 2014b, \apj, 791, 91

%\bibitem[Clanton \& Gaudi(2016)]{clanton16}Clanton, C.D. \& Gaudi, B.S. 2016, \apj, 819, 125

%\bibitem[Claret(2000)]{claret00}Claret, A.\ 2000, \aap, 363,1081

\bibitem[DePoy et al.(2003)]{depoy03} DePoy, D.L., Atwood, B., Belville, S.R.,
 et al.\ 2003, SPIE 4841, 827

%\bibitem[Dong et al.(2006)]{ob04343} Dong, S., DePoy, D.L., Gaudi, B.S., et al. 2006, \apj, 642, 842

%\bibitem[Dong et al.(2009)]{ob05071b} Dong, S., Gould, A., Udalski, A., et al. 2009, \apj, 695, 970

%\bibitem[Dong et al.(2009)]{mb07400} Dong, S., Bond, I.A., Gould, A., et al. 2009, \apj, 695, 970

%\bibitem[Dominik(1999)]{dominik99} Dominik, M. 1999, \aap, 349, 108

%\bibitem[Dominik et al.(2007)]{signalmen} Dominik, M., Rattenbury, N.J., Allan, A., et al. 2007, \mnras, 380, 792.

\bibitem[Duquennoy \& Mayor(1991)]{dm91}Duquennoy, A., \& Mayor, M. 1991, \aap, 248, 485

%\bibitem[Evans et al.(2016)]{evans16} Evans, D.F., Southworth, J., Maxted, P.F.L., et al. 2016, \aap 589, 58.

%\bibitem[Gaudi(1998)]{gaudi98} Gaudi, B.S.\ 1998, \apj, 506, 533

%\bibitem[Gaudi(2012)]{gaudi12} Gaudi, B.S.\ 2012, \araa, 50, 411

%\bibitem[Gaudi \& Gould(1997)]{gaudi97} Gaudi, B.S. \& Gould, A.\ 1997, \apj, 477, 152

%\bibitem[Gaudi \& Gould(1997)]{gaudi97} Gaudi, B.S. \& Gould, A.\ 1997, \apj, 486, 85

%\bibitem[Gaudi \& Sackett(2000)]{gaudi00} Gaudi, B.S. \& Sackett, P. 2000, \apj, 528, 56

%\bibitem[Gaudi et al.(2002)]{gaudi02} Gaudi, B.S., Albrow, M.D., An, J.\ 2002, \apj, 566, 463

\bibitem[Gould(1992)]{gould92} Gould, A. 1992, \apj, 392, 442

%\bibitem[Gould(1994a)]{gould94a} Gould, A. 1994a, \apjl, 421, L71

%\bibitem[Gould(1994b)]{gould94b} Gould, A. 1994b, \apjl, 421, L75

%\bibitem[Gould(1995)]{gould95} Gould, A. 1995, \apjl, 441, L21

\bibitem[Gould(1995)]{gould95} Gould, A. 1995, \apj, 446, L71

%\bibitem[Gould(1996)]{gould96} Gould, A. 1996, \apj, 470, 201

%\bibitem[Gould(1997)]{gould97} Gould, A. 1997, The Hollywood Strategy for Microlensing Detection of Planets, in Variables Stars and the Astrophysical Returns of the Microlensing Surveys. Eds. R. Ferlet, J.-P. Maillard and B. Raban. Gif-sur-Yvette, France : Editions Frontieres, p.125

\bibitem[Gould(2000)]{gould00} Gould, A. 2000, \apj, 542, 785

%\bibitem[Gould(2004)]{gould04} Gould, A. 2004, \apjl, 606, 319

%\bibitem[Gould(2008)]{gould08} Gould, A.\ 2008, \apj, 681, 1593

%\bibitem[Gould \& Andronov(1999)]{gouldandronov99} Gould, A.\ \& Andronov, N. 1999, \apj, 516, 236


%\bibitem[Gould \& Gaucherel(1997)]{gg97} Gould, A. \& Gaucherel. 1997, \apj, 477, 580

%\bibitem[Gould \& Horne(2013)]{gouldhorne} Gould, A. \& Horne, K. 2013, \apjl, 779, L28

\bibitem[Gould \& Loeb(1992)]{gouldloeb} Gould, A. \& Loeb, A. 1992, \apj, 396, 104

%\bibitem[Gould \& Yee(2012)]{gould12} Gould, A. \& Yee, J.C. 2012, \apj, 755, L17

%\bibitem[Gould et al.(2006)]{ob05169} Gould, A., Udalski, A., An, D.\ et al.\ 2006, \apj, 644, L37

%\bibitem[Gould et al.(2009)]{ob07224} Gould, A., Udalski, A., Monard, B.\ et al.\ 2013, \apj, 698, L147

%\bibitem[Gould et al.(2014)]{ob130341} Gould, A., Udalski, A., Shin, I.-G.\ et al.\ 2014, Science, 345, 46


%\bibitem[Gould et al.(2013)]{prop2013} Gould, A., Carey, S., \& Yee, J. 2013, 2013spitz.prop.10036

%\bibitem[Gould et al.(2014)]{prop2014} Gould, A., Carey, S., \& Yee, J. 2014, 2014spitz.prop.11006

%\bibitem[Gould et al.(2015a)]{prop2015a} Gould, A., Yee, J., \& Carey, S., 2015a, 2015spitz.prop.12013

%\bibitem[Gould et al.(2015b)]{prop2015b} Gould, A., Yee, J., \& Carey, S., 2015b, 2015spitz.prop.12015

%\bibitem[Gould et al.(2016)]{prop2016} Gould, A., Yee, J., \& Carey, S., 2016, 2015spitz.prop.13005

\bibitem[Gould et al.(2010)]{gould10} Gould, A., Dong, S., Gaudi, B.S.\ et al.\ 2010, \apj, 720, 1073

%\bibitem[Gould et al.(2013)]{gould13} Gould, A., Shin, I.-G., Han, C. et al.\ 2013, \apj, 768, 126

%\bibitem[Grether \& Lineweaver(2006)]{grether06} Grether, D., \& Lineweaver, C.H., 2006, \apj, 640, 1051

\bibitem[Groenewegen(2004)]{groenewegen04} Groenewegen, M.A.T., 2004, \mnras, 353, 903

%\bibitem[Griest \& Safizadeh(1998)]{griest98} Griest, K.\ \& Safizadeh, N.\ 1998, \apj, 500, 37

%\bibitem[Han \& Gould(1995)]{han95} Han, C. \& Gould, A.\ 1995, \apj, 447, 53

%\bibitem[Han \& Gould(2003)]{han03} Han, C. \& Gould, A.\ 2003, \apj, 592, 172

%\bibitem[Han(2006)]{han06} Han, C.  2006, \apj, 638, 1080

%\bibitem[Han et al.(2016)]{ob150768}Han, C., Udalski, A., Lee, C.-U., et al. 2016, \apj, 827, 11

%\bibitem[Henderson et al.(2016)]{henderson16} Henderson, C.B., Poleski, R., Penny, M. et al. 2016 \pasp 128, 124401

%\bibitem[Hodgkin et al.(2009)]{Hodgkin.2009.A} Hodgkin, S.~T., Irwin, M.~J., Hewett, P.~C., \& Warren, S.~J.\ 2009, \mnras, 394, 675

%\bibitem[Hwang et al.(2018)]{ob170173} Hwang, K.-H., Udalski, A., Shvartzvald, Y. et al. 2018, \aj, 155, 20 %, arXiv:1709.08476

%\bibitem[Irwin et al.(2004)]{Irwin.2004.A} Irwin, M.~J., Lewis, J., Hodgkin, S., et al.\ 2004, \procspie, 5493, 411

%\bibitem[Jung et al.(2014)]{jung14}Jung, Y.\ K., Park, H., Han, C., et al. 2014 \apj, 786, 85

%\bibitem[Jung et al.(2015)]{jung15}Jung, Y.\ K., Udalski, A., Sumi, T., et al.\ 2015 \apj, 798, 123

%\bibitem[Jung et al.(2013)]{jung13}Jung, Y.\ K., Han, C., Gould, A., \& Maoz, D. 2013 \apjl, 768, L7

%\bibitem[Kayser et al.(1986)]{kayser86} Kayser, R., Refsdal, S., \& Stabell, R. 1986, \aap, 166, 36

%\bibitem[Kennedy \& Kenyon(2008)]{kennedy08} Kennedy, G.M. \& Kenyon, S.J. 2008, \apj, 673, 502

%\bibitem[Kervella et al.(2004)]{kervella04} Kervella, P., Th{\'e}venin, F., Di Folco, E., \& S{\'e}gransan, D.\ 2004, \aap, 426, 297

\bibitem[Kim et al.(2016)]{kmtnet} Kim, S.-L., Lee, C.-U., Park, B.-G., et al.  2016, JKAS, 49, 37

\bibitem[Kim et al.(2018a)]{eventfinder} Kim, D.-J., Kim,  H.-W., Hwang, K.-H., et al., 2018a, \aj, 155, 76

\bibitem[Koshimoto et al.(2014)]{ob08355}Koshimoto1, N., Udalski2,  A., Sumi, T., et al. 2014, \aj, 788, 128

%\bibitem[Kubas et al.(2012)]{mb07192k}Kubas, D., Beaulieu, J.-P., Bennett, D.P., et al. 2012, \aap, 540A, 78

%\bibitem[Marcy \& Butler(2000)]{marcy00} Marcy, G.W. \& Butler, R.P. 2000, \pasp, 112,137

%\bibitem[Mao \& Paczy\'nski(1991)]{mao91} Mao, S.\ \& Paczy\'nski, B.\ 1991, \apj, 374, 37

\bibitem[Minniti et al.(2010)]{vvv-survey1}Minniti, D., Lucas, P. W., Emerson, J. P., et al. 2010, New Astron., 15, 433

\bibitem[Minniti(2018)]{vvvx-survey}Minniti, D.\ 2018, in The Vatican Observatory,  Castel Gandolfo: 80th  Anniversary Celebration (ed. G. Gionti, S.J., \&  J.-B. Kikwaya Eluo, S.J). Astrophysics and Space Science Proceedings, 51, 63

%\bibitem[Mr\'oz et al.(2017)]{ob160596}Mr\'oz, P., Han, C., Udalski, A. et al.\ 2017, \aj, 153, 143

%\bibitem[Muraki et al.(2011)]{mb09266} Muraki, Y., Han, C., Bennett, D.P., et al.\ 2011, \apj, 741, 22

\bibitem[Nataf et al.(2013)]{nataf13} Nataf, D.M., Gould, A., Fouqu\'e, P. et al. 2013, \apj, 769, 88

\bibitem[Navarro et al.(2017)]{vvv-ulens1} Navarro, M.G., Minniti, D., \& Contreras-Ramos, R. 2017, \apj, 851, L13

\bibitem[Navarro et al.(2018)]{vvv-ulens2} Navarro, M.G., Minniti, D., \& Contreras-Ramos, R. 2018, \apj, 865, L5

\bibitem[Navarro et al.(2019)]{vvv-ulens3} Navarro, M.G., Minniti, D., \& Contreras-Ramos, R. 2019, submitted

%\bibitem[Pascucci et al.(2018)]{kepq}Pascucci, I., Mulders, G.D., Gould, A., \& Fernandes, R. 2018, \apj, submitted

\bibitem[Paczy\'nski(1986)]{pac86} Paczy\'nski, B.\ 1986, \apj, 304, 1

%\bibitem[Park et al.(2013)]{park13}Park, H., Udalski, A., Han, C., et al.\ 2013, \apj, 778, 134

%\bibitem[Park et al.(2015)]{park15}Park, H., Udalski, A., Han, C., et al.\ 2015, \apj, 805, 117

%\bibitem[Pejcha \& Heyrovsk\'y(2009)]{pejcha09} Pejcha, O., \& Heyrovsk\'y, D.\ 2009, \apj, 690, 1772

%\bibitem[Penny et al.(2016)]{penny16} Penny, M.T., Henderson, C.B., \& Clanton, C.\ 2016, \apj, 830, 150

\bibitem[Poleski et al.(2014a)]{ob08092} Poleski, R., Skowron, J., Udalski, A.,  et al.\ 2014a, \apj, 755, 42

\bibitem[Poleski(2016)]{poleski16} Poleski, R.\ 2016, \mnras, 455, 3656

%\bibitem[Poleski et al.(2014b)]{ob120406} Poleski, R., Udalski, A., Dong, S.\ et al.\ 2014b, \apj, 782, 47

%\bibitem[Poleski et al.(2014b)]{mb12006} Poleski, R., Udalski, A., Bond, I.A..\ et al.\ 2017, arXiv:1704.01121

%\bibitem[Refsdal(1966)]{refsdal66} Refsdal, S. 1966, \mnras, 134, 315

\bibitem[Rattenbury et al.(2015)]{mb10353}Rattenbury, N.J., Bennett, D.P., Sumi, T., et al.\ 2015, \mnras, 454, 946

%\bibitem[Rhie et al.(2000)]{rhie00} Rhie, S.H., Bennett, D.P., Becker, A.C. et al. 2000, ApJ, 533, 378

%\bibitem[Ryu et al.(2017)]{ob160693} Ryu, Y.-H., Udalski, A., Yee, J.C. et al. 2017, \aj, 154, 247 %arXiv:1707.12222

\bibitem[Saito et al.(2012)]{vvv-survey2}Saito, R.K., Hempel, M., Minniti, D., et al. 2012, \aap, 537, A107

%\bibitem[Santerne et al.(2016)]{santerne16} Santerne, A., Beaulieu, J.-P., Rojas Ayala, B., et al. 2016, \aap, in press  arXiv:1610.04446


%\bibitem[Schechter et al.(1993)]{dophot} Schechter, P.L., Mateo, M., \& Saha, A. 1993, \pasp, 105, 1342

%\bibitem[Schmidt(1968)]{schmidt68} Schmidt, M., 1968 \apj, 151, 393

%\bibitem[Schneider \& Weiss(1988)]{schneider88}Schneider, P., \& Weiss, A. 1988, \apj, 330, 1

%\bibitem[Shin et al.(2012a)]{ob110417}Shin, I.-G., Han, C., Choi, J.-Y., et al.\ 2012a, \apj, 755, 91

%\bibitem[Shin et al.(2012b)]{shin12b}Shin, I.-G., Han, C., Gould, A., et al.\ 2012b, \apj, 760, 116


%\bibitem[Shin et al.(2016)]{ob150954} Shin, I.-G., Ryu, Y.H, Udalski, A.\ et al. 2016, JKAS, 49, 73

\bibitem[Shin et al.(2018)]{ob161045} Shin, I.-G., Yee, J.C., Skowron, J.\ et al. 2018, \apj, 863, 23

%\bibitem[Shvartzvald et al.(2014)]{mb11322} Shvartzvald, Y., Maoz, D., Kaspi, S.\ et al.\ 2014, \mnras, 439, 604

\bibitem[Shvartzvald et al.(2016)]{shvartzvald16} Shvartzvald, Y., Maoz, D., Udalski, A.\ et al.\ 2016, \mnras, 457, 4089

\bibitem[Shvartzvald et al.(2017)]{ukirt-5events} Shvartzvald, Y.,Bryden, G., Gould, A., et al.\ 2017, \aj, 153, 61

\bibitem[Shvartzvald et al.(2018)]{ub17001} Shvartzvald, Y., Calchi Novati, S., Gaudi, B.S., et al.\ 2018, \apj, 857, 8


%\bibitem[Shvartzvald et al.(2015)]{ob151285} Shvartzvald, Y., Udalski, A., Gould, A.\ et al.\ 2015, \apj, 814, 111

%\bibitem[Shvartzvald et al.(2017)]{ob161195b} Shvartzvald, Y., Yee, J.C., Calchi Novati, S.\ et al.\ 2017, \apjl, 840, L3

%\bibitem[Skottfelt et al.(2015)]{skottfelt15} Skottfelt, J., Bramich, D. M., Hundertmark, M., et al., 2015, \aap, 574, 54



\bibitem[Skowron et al.(2011)]{ob09020}Skowron, J., Udalski, A., Gould, A et al.\ 2011, \apj, 738, 87

%\bibitem[Skowron et al.(2011)]{mb11028} Skowron, J., Udalski, A., Poleski, R. et al.\ 2016, \apj, 820, 4

%\bibitem[Skowron et al.(2016)]{skowron16} Skowron, J., Udalski, A., Koz{\l}owski, S. et al.\ 2016, Acta Astron., 66, 1

%\bibitem[Spergel et al.(2013)]{spergel13} Spergel, D.N., Gehrels, N., Breckinridge, J., et al. 2013, arXiv:1305.5422

%\bibitem[Smith et al.(2003)Smith, Mao \& Paczy\'nski]{smp03} Smith, M., Mao, S., \& Paczy\'nski, B., 2003, \mnras, 339, 925

%\bibitem[Street et al.(2013)]{street13}Street, R.\ A., Choi, J.-Y., Tsapras, Y., et al.\ 2013, \apj, 763, 67

%\bibitem[Street et al.(2016)]{ob150966} Street, R., Udalski, A., Calchi Novati, S.\ et al.\ 2016, \apj, 829, 93.

%\bibitem[Sumi et al.(2010)]{ob07368} Sumi, T., Bennett, D.P., Bond, I.A., et al.\ 2010, \apj, 710, 1641

%\bibitem[Sumi et al.(2016)]{mb13605} Sumi, T., Udalski, A., Bennett, D.P., et al.\ 2016 \apj, in press arXiv:1512.00134

\bibitem[Suzuki et al.(2014)]{mb08379} Suzuki, D., Udalski, A., Sumi, T., et al. 2014, \apj, 780, 123

%\bibitem[Suzuki et al.(2016)]{suzuki16} Suzuki, D., Bennett, D.P., Sumi, T., et al. 2016, \apj, 833, 145

%\bibitem[Thompson(2013)]{thompson13} Thompson, T.A. 2013, \mnras, 431, 63

%\bibitem[Tsapras et al.(2014)]{ob120406b}Tsapras, Y., Choi, J.-Y., Street, R.-A., et al. 2014, \apj, 782, 48

%\bibitem[Udalski(2003)]{ews2} Udalski, A. 2003, Acta Astron., 53, 291

%\bibitem[Udalski et al.(1994)]{ews1} Udalski, A.,Szymanski, M., Kaluzny, J., Kubiak, M., Mateo, M.,  Krzeminski, W., \& Paczy\'nski, B. 1994, Acta Astron., 44, 227

\bibitem[Udalski et al.(2015a)]{ogleref} Udalski, A., Syzmanski, M.K., \& Szymanski, G., et al. 2015a, AcA, 65, 1

%\bibitem[Udalski et al.(2005)]{ob05071} Udalski, A., Jaroszy\'nski, M., Paczy\'nski, B, et al. 2005, \apj, 628, L109.

%\bibitem[Udalski et al.(2015)]{ob140124} Udalski, A., Yee, J.C., Gould, A., et al. 2015, \apj, 799, 237

%\bibitem[Wambsganss(1997)]{wambs97}Wambsganss, J. 1997, \mnras, 284, 172

\bibitem[Wo\'zniak(2000)]{wozniak2000} Wo\'zniak, P.~R. 2000, Acta Astron., 50, 421

%\bibitem[Wyrzykowski et al.(2015)]{wyr15}Wyrzykowski, \L., Rynkiewicz, A.E., Skowron, J., et al. 2015, \apjs, 216, 12

%\bibitem[Udalski et al.(2015b)]{ogleiv} Udalski, A., Szyma\'nski, M.K. \& Szyma\'nski, G. 2015b, Acta Astronom., 65, 1

%\bibitem[Yee et al.(2012)]{mb11293} Yee, J.C., Shvartzvald, Y., Gal-Yam, A.\ et al.\ 2012, \apj, 755, 102

%\bibitem[Yee et al.(2013)]{mb10311} Yee, J.C., Hung, L.-W., Bond, I.A.\ et al.\ 2013, \apj, 769, 77

%\bibitem[Yee et al.(2015)]{yee15} Yee, J.C., Gould, A., Beichman, C., 2015, \apj, 810, 155

%\bibitem[Yee et al.(2016)]{yee16} Yee, J.C., Johnson, J.A., Skowron, J., et al. 2016, \apj, 821, 121

\bibitem[Yoo et al.(2004)]{ob03262} Yoo, J., DePoy, D.L., Gal-Yam, A.\ et al.\ 2004, \apj, 603, 139

%\bibitem[Zhu et al.(2014)]{Zhu:2014} Zhu, W., Penny, M., Mao, S., Gould, A., \& Gendron, R.\ 2014, \apj, 788, 73

%\bibitem[Zhu et al.(2015)]{zhu15} Zhu, W., Gould, A., XXX, et al. 2015, \apj, 814, 129

%\bibitem[Zhu et al.(2015)]{ob141050} Zhu, W., Udalski, A., Gould, A., et al. 2015, \apj, 805, 8

%\bibitem[Zhu et al.(2016)]{ob150763} Zhu, W., Calchi Novati, S., Gould, A., et al. 2016, \apj, 825, 60

%\bibitem[Zhu et al.(2017)]{zhu17} Zhu, W., Udalski, A., Calchi Novati, S.,  et al. 2017, \aj, 154, 210 %arXiv:1701.05191

\end{thebibliography}
\end{document}